\documentstyle[aps,epsfig]{revtex}

\begin{document}

\title{Quantum cryptography using photon source based on postselection
from entangled two-photon states}

\author{Jan Pe\v{r}ina, Jr.$ {}^{a,b} $, \\
Ond\v{r}ej Haderka$ {}^{a,b} $, \\
Jan Soubusta$ {}^{a,b} $, \\
$ {}^{a} $ Joint Laboratory of Optics
of Palack\'{y} University and \\
Institute of Physics of Academy of Sciences of the Czech
Republic, \\
17. listopadu 50A, 772 00 Olomouc,
Czech Republic \\
$ {}^{b} $ Department of Optics, Faculty of Natural Sciences, Palack\'{y}
University, \\ 17. listopadu 50,
772 00 Olomouc, Czech Republic }

\maketitle

\begin{abstract}
A photon source based on postselection from
entangled photon pairs produced by parametric frequency down-conversion is
suggested. Its ability to provide good approximations of single-photon
states is examined.
Application of this source in quantum cryptography for quantum key
distribution is discussed. Advantages of the source compared to
other currently used sources are clarified.
Future prospects of the photon source are outlined.
\end{abstract}



\section{Introduction}

Entangled photon pairs produced by spontaneous parametric frequency
down-conversion \cite{mandelwolf,milburnwalls} have recently been widely used
in experimental quantum physics. They have been successfully applied
in the research of fundamental problems of quantum theory \cite{perina}.
Among others, a direct application of nonclassical properties of such
states in optical communications has been suggested \cite{ekert}
in 1991.

Since then the area of quantum communications underwent an immense progress.
Quantum key distribution (QKD) became a well understood scheme for establishing
a provably secure shared secret not only at a theoretical level.
Experimental realization brought QKD at the disposal of future commercial
applications \cite{bruss}.
Most of the practical QKD schemes designed until now relied on dim coherent
pulses as a carrier of qubits. While this scheme suffers from the lack of the
ultimate proof of security \cite{mayers,lochau}, its security is very well
defined and understood \cite{lutkenhaus}.

Recently, the idea to use correlated photon pairs for QKD has been revisited
in two different ways. First, the laboratory realization of the original Ekert's
protocol has been improved and modified
\cite{tittel,jennewein,sergienko,ribordy} and
its security has been addressed \cite{naik}.
A passive scheme for choosing from two possible transmission
bases has been suggested and realized. Possibility
of multiphoton attacks on QKD is substantially reduced
in this scheme.
Second, the fact that down-converted
photons are always produced in pairs has been used to suggest a new source of
photons applicable in quantum cryptography \cite{lutkenhaus,brassard,greece}.
The state describing such correlated fields cannot be factorized into a product
of states of signal and idler beams. When a measurement is performed on one
of the beams, the whole state including the other beam is changed.
When a photon is detected in, e.g., the signal beam, we
know that its twin must be present
in the idler beam. This suggests to construct a single-photon source as
follows:
Perform a photon-number measurement on one of the beams and select only those
cases when a single photon has been detected. Then
there is a single photon in the other beam with a high
probability
and this photon is used for cryptography.
Filtering of both vacuum and multiphoton states contributes to the
security of QKD. Moreover, as is shown in the
paper, this scheme provides higher values of transmission rates.
A passive scheme that lowers the vulnerability to multiphoton attacks
may also be implemented in this case.

Practical existence of a one-photon source would
help formulating general security proofs \cite{mayers} of quantum key
distribution protocols in secure quantum cryptography.
It would also make practical schemes more efficient \cite{lutkenhaus}.

A realistic model of such source of photons including
imperfections encountered in the laboratory is developed in the
paper. Efficiency of the
postselection procedure as well as applicability of the source in real
quantum cryptography are discussed.

\section{Model of the source}

We assume that a postselection device is placed in the signal beam. This device
yields a simple yes-no result (a trigger) and, based on this result, the state
in the idler beam is either coupled to the transmission line or rejected. It
is, however, not an easy task to construct a practical photon-number measuring
device. Generally used photon-counting detectors (avalanche photodiodes or
photomultipliers) use many-order noisy amplification processes that smear out
resolution of small photon numbers. In our work we use a model of a photon
number measuring device based on $ 1\times N $ coupler \cite{jex}.
We note, that novel detectors capable to resolve small numbers of
photons and sources of single photons occured recently \cite{kim}. However,
they work only at very low temperatures and having practical QKD on mind,
we do not consider them here.
Performance of a measuring device based on $ 1\times N $
coupler and $ N $ detectors and a photon-number resolving
detector in the preparation of a state in postselection procedure
has been studied in \cite{KokB}.

In our model \cite{greece,nase} we assume that the down-conversion process
is pumped by either cw or pulsed laser beam. The signal and idler beams are
selected by filters and pinholes (geometrically and spectrally filtered). The
filtering is in general imperfect, i.e. sometimes only one of the members of
the pair reaches a detector. Detectors have limited quantum
efficiencies and
they are not capable of resolving the photon number, they just click in the
presence of the signal containing one or more photons. In addition, there are
noise detections coming both from dark counts of the detectors and from
stray light in the laboratory. The scheme including all these imperfections
is given in Fig.~1.
\begin{figure}    
 \begin{center}
   \epsfig{file=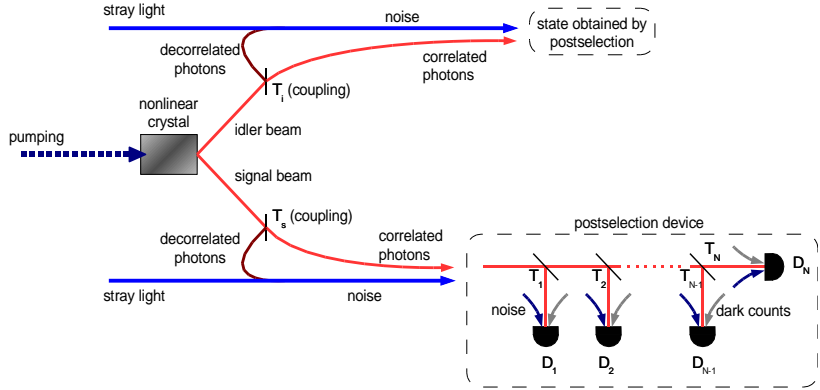,angle=0,width=0.6\hsize}
 \end{center}
 \caption{Scheme of the model. Photons of the pump beam are
split in the nonlinear crystal to pairs of mutually entangled photons.
The signal beam is then coupled to a postselection device consisting of
a $ 1\times N $
coupler and $ N $ detectors ($ T_i $ stands for the intensity
transmission coefficient to $ i $th detector, $ T_i = |t_i|^2 $).
Coupling is imperfect and decorrelated
photons contribute to the noise impinging on the detectors that exhibit also
internal dark-count noise. The idler beam suffers similar coupling
problems.}
\end{figure}

Detection of a photoelectron is described by the following
projection operator ($ \eta $ is quantum efficiency of the
detector):
\begin{equation}           
\label{Eclick}
 \hat{P}^{\rm det}=\sum _{n=1}^{\infty }\left[ 1-
 \left( 1-\eta \right) ^{n}\right]
 \left| n\right\rangle \left\langle n\right| +d\!
 \sum^{\infty }_{n=0}\left( 1-\eta \right) ^{n}\left| n\right\rangle
 \left\langle n\right| ,
\end{equation}
where $ d $ represents a total noise count rate determined as
\begin{equation}          
\label{d}
 d=d^{\rm dark}+\left( 1-d^{\rm dark}\right) \! d^{\rm noise},
\end{equation}
when both dark counts and noise coming from stray light and
decorrelated photons are taken
into account (for details, see Appendix A).
The projection operator $ \hat{P}^{\rm nodet} $ appropriate
in the case when a photoelectron
does not occur in the detector has the form
($ \hat{P}^{\rm nodet} = \hat{1} - \hat{P}^{\rm det} $):
\begin{equation}         
\label{Enoclick}
 \hat{P}^{\rm nodet}=\left( 1-d\right) \! \sum ^{\infty }_{n=0}
 \left( 1-\eta \right) ^{n}\left| n\right\rangle \left\langle n\right| .
\end{equation}

The light field emerging from the output of the nonlinear crystal
is in an entangled multimode state. However, it can be described
as an entangled state of two effective modes (one for the signal field,
the other for the idler field) by the statistical operator
$ \hat{\rho}_{S,I} $:
\begin{eqnarray}     
 \hat{\rho}_{SI} &=& |\psi\rangle \langle \psi | , \nonumber \\
 | \psi \rangle &=& \sum_{n=0}^{\infty } c_{n}
 | n,n\rangle _{S,I}\! ,
\label{Psi}
\end{eqnarray}
where the indices $ S $ and $ I $ refer to the signal and idler beams,
respectively. As is shown in Appendix B, the statistics of pairs of
photons in two effective modes is Poissonian, i.e.:
\begin{equation}     
\label{Poisson}
\left| c_{n}\right| ^{2}=\frac{\mu ^{n}}{n!}e^{-\mu },
\end{equation}
$ \mu $ being the mean number of pairs generated during a detection interval.

Both the signal and idler beams experience losses before they are
detected (spectral filtering by interference filters, geometrical
filtering by pinholes and other elements in the experimental setup).
We represent all these losses by quantally described beamsplitters
\cite{campos};
a beamsplitter in the signal (idler) field has a transmission
coefficient $ T_{S} $ ($ T_{I} $). Diagonal elements of the
statistical operator $ \hat{\rho}'_{S,I} $
in the Fock-state basis then have the form
(the signal and idler fields are partially decorrelated;
$ R_S = 1 - T_S $, $ R_I = 1 - T_I $):
\begin{equation}       
 (\hat{\rho'}_{S,I})_{l_S l_I,l_S l_I}=
 \sum_{n={\rm max}(l_S,l_I)}^{\infty} |c_n|^2
 \left( \begin{array}{c} n \\ l_S \end{array} \right)
 T_S^{l_S} R_S^{n-l_S}
 \left( \begin{array}{c} n \\ l_I \end{array} \right)
 T_I^{l_I} R_I^{n-l_I} ,
\end{equation}
where the symbol $ {\rm max} $ denotes the maximum function.
We limit ourselves only to the determination of diagonal elements
of the statistical operator $ \hat{\rho'}_{S,I} $, because they
are sufficient for the description of the detection process.

Photon-number measurement in the signal field may be
approximatelly reached using a $ 1\times N $
coupler and $ N $ detectors. Provided that the mean photon number
of the signal field is much lower that the number of detectors
and the detectors exhibit moderate quantum efficiencies and
dark count rates, only one detector detects a photon on
single-photon signal
while multiple detections occur on multi-photon signals with high probability.
A $ 1\times N $ coupler is described by the unitary transformation,
\begin{equation}     
 \hat{a'}_S = \sum_{j=1}^{N} t_j \hat{a}_{j} ,
\end{equation}
where $ \hat{a'}_S $ stands for the annihilation operator
at the input of the coupler, $ \hat{a}_j $ is the annihilation
operator at the $ j $-th output of the coupler, and
$ t_j $ means the amplitude transmission coefficient
of a photon propagating from the input to the $ j $-th output
of the coupler ($ j=1,\ldots,N $).

The statistical operator $ \hat{\rho}^{\rm post}_{I,k} $
describing the idler field after a signal photon has been
detected at the $ k $-th detector and no photon has been
detected at all other detectors beyond the $ 1\times N $
coupler is determined as follows:
\begin{equation}     
 \hat{\rho}^{\rm post}_{I,k} = \frac{
 {\rm Tr}_{S} (\hat{\rho'}_{S,I} \hat{P}^{\rm det}_{k}
 \prod_{j=1,\ldots,N;j\ne k} \hat{P}^{\rm nodet}_{j} ) }{
 {\rm Tr}_{S,I} (\hat{\rho'}_{S,I} \hat{P}^{\rm det}_{k}
 \prod_{j=1,\ldots,N;j\ne k} \hat{P}^{\rm nodet}_{j} ) },
\end{equation}
where the projection operators $ \hat{P}^{\rm det}_j $ given
in Eq.~(1)
and $ \hat{P}^{\rm nodet}_j $ defined in Eq.~(3)
are related to the $ j $-th
detector. Using the statistical operator $ \hat{\rho'}_{S,I} $
given in Eq.~(6) together with the relation appropriate
for the $ 1\times N $ coupler in Eq.~(7), we arrive at the
expression:
\begin{eqnarray}     
 (\hat{\rho}^{\rm post}_{I,k})_{l_I,l_I} &=& \frac{1}{r_{I,k}}
 \left\{ \left[ \prod_{l=1,\ldots,N; l\ne k} (1-d_l) \right]
 \sum_{n=l_I}^{\infty} |c_n|^2 \left( \begin{array}{c}
 n \\ l_I \end{array} \right) T_I^{l_I} R_I^{n-l_I} {\cal A}_k^n
 \right. \nonumber \\
 & & \mbox{} - \left. \left[ \prod_{l=1}^{N} (1-d_l) \right]
 \sum_{n=l_I}^{\infty} |c_n|^2 \left( \begin{array}{c}
  n \\ l_I \end{array} \right) T_I^{l_I} R_I^{n-l_I} {\cal B}^n
 \right\} ,
\end{eqnarray}
and
\begin{eqnarray}   
 {\cal A}_k &=& R_S + T_S \left( |t_k|^2 + \sum_{l=1,\ldots,N;
 l\ne k} |t_l|^2 (1-\eta_l) \right) , \nonumber \\
 {\cal B} &=& R_S + T_S \left( \sum_{l=1}^{N} |t_l|^2 (1-\eta_l) \right)
 .
\end{eqnarray}
The symbol $ \eta_j $ stands for the quantum efficiency of
the $ j $-th detector. The normalization constant $ r_{I,k} $
is determined as follows:
\begin{equation}  
 r_{I,k} = \left[ \prod_{l=1,\ldots,N; l\ne k} (1-d_l) \right]
 \sum_{n=0}^{\infty} |c_n|^2 {\cal A}_k^n
  - \left[ \prod_{l=1}^{N} (1-d_l) \right]
 \sum_{n=0}^{\infty} |c_n|^2 {\cal B}^n .
\end{equation}

Noise in the signal beam coming from both stray light and
decorrelated photons may be included into the model through
the constants $ d^{\rm noise} $ in Eq.~(2). The influence of noise
in the idler beam has to be described more precisely.
We consider a chaotic field with the statistical operator
$ \hat{\rho}_{I,k}^{\rm res} $ (for details, see Appendix~B)
statistically independent on the
idler field stemming from the postselection procedure.
The statistical operator $ \hat{\rho}^{\rm mix}_{I,k} $ of
the overall field at the detector in the idler beam is
given as follows \cite{perinabook}:
\begin{equation}   
 \hat{\rho}_{I,k}^{\rm mix} = \sum_{n=0}^{\infty} | n \rangle_I
 {}_I \langle n| \sum_{m=0}^{n} (\hat{\rho}_{I,k}^{\rm post})_{m,m}
 (\hat{\rho}_{I,k}^{\rm res})_{n-m,n-m} .
\end{equation}

We now consider Poissonian statistics of the generated pairs
of photons (the coefficients $ c_n $ are given in Eq.~(5))
and chaotic noisy fields both in the signal and idler
beams (see Appendix~B):
\begin{eqnarray}   
 (\hat{\rho}_{I,k}^{\rm res})_{n,n} &=& (1-\nu^{\rm res}_{I,k})
 (\nu^{\rm res}_{I,k})^n , \hspace{0.5cm}
  \nu^{\rm res}_{I,k} = \frac{\mu^{\rm res}_{I,k}}{1+
   \mu^{\rm res}_{I,k}} , \nonumber \\
 (\hat{\rho}_{S,j}^{\rm res})_{n,n} &=& (1-\nu^{\rm res}_{S,j})
 (\nu^{\rm res}_{S,j})^n , \hspace{0.5cm}
  \nu^{\rm res}_{S,j} = \frac{\mu^{\rm res}_{S,j}}{
  1+ \mu^{\rm res}_{S,j}} , \hspace{0.5cm} j=1,\ldots,N .
\end{eqnarray}
The symbol $ \mu_{I,k}^{\rm res} $ denotes the mean number of
noisy photons in the idler beam and $ \mu^{\rm res}_{S,j} $
is the mean number of noisy photons in the signal field
at the $ j $-th detector.
The diagonal matrix elements of the statistical operator
$ \hat{\rho}_{I,k}^{\rm mix} $ given in Eq.~(12) then take
the form:
\begin{eqnarray}    
 (\hat{\rho}_{I,k}^{\rm mix})_{nn} &=& \frac{ 1-
 \nu^{\rm res}_{I,k} }{
 \exp(-\mu {\cal B}) - (1-d_k) \exp(-\mu {\cal A}_k ) }
  \nonumber \\
  & & \mbox{} \times \left\{ \exp (-\mu {\cal B})
 \exp(-\mu {\cal A}_k T_I) (\nu_{I,k}^{\rm res})^n
  f_n\left( \frac{\mu {\cal A}_k T_I}{\nu_{I,k}^{\rm res}} \right)
 \right. \nonumber \\
  & & \left. \mbox{} - (1-d_k) \left[ \exp(-\mu {\cal A}_k)
 \exp(-\mu {\cal B} T_I) (\nu_{I,k}^{\rm res})^n
  f_n\left( \frac{\mu {\cal B} T_I}{\nu_{I,k}^{\rm res}} \right)
  \right] \right\} ,
  \nonumber \\
 & &
\end{eqnarray}
and
\begin{equation}    
 f_n(x) = \sum_{l=0}^{n} \frac{x^l}{l!} .
\end{equation}
The normalization constant $ r_{I,k} $ is determined according
to:
\begin{equation}  
 r_{I,k} = \left[ \prod_{l=1,\ldots,N; l\ne k} (1-d_l) \right]
  \exp[-\mu(1-{\cal A}_k)]
  - \left[ \prod_{l=1}^{N} (1-d_l) \right]
   \exp[-\mu(1-{\cal B})] .
\end{equation}

The noisy field in the signal beam is
given by the photons that lost their twins (the term $ T_S \mu $
in Eq.~(17) below, see
Appendix~C for details) and by additional noisy photons
coming, e.g., from stray light
(the mean number of additional noisy photons is denoted as
$ \mu^{\rm res, add}_S $). We then have:
\begin{equation}   
 \mu^{\rm res}_{S,j} = |t_j|^2 \mu^{\rm res}_S ; \hspace{1cm}
 \mu^{\rm res}_S = (T_S \mu + \mu^{\rm res, add}_S) .
\end{equation}
The constants $ d^{{\rm noise}}_j $ in Eq.~(2) describing the influence
of noise in the signal beam at the $ j $-th detector are then
determined as follows:
\begin{equation}    
 d^{\rm noise}_j = \frac{ \eta_j \mu^{\rm res}_{S,j} }{ 1+ \eta_j
 \mu^{\rm res}_{S,j} } .
\end{equation}
Similarly, assuming that the noisy field in the idler beam
consists of both idler photons without
their twins in the signal field and additional noisy photons
with the mean number of photons denoted as
$ \mu^{\rm res, add}_{I} $, we have:
\begin{equation}    
 \mu^{\rm res}_{I,k} = T_I \mu + \mu^{\rm res, add}_I.
\end{equation}

If narrow spectra of the down-converted fields are considered,
the statistics of generated pairs is given by the Bose-Einstein
distribution \cite{milburnwalls}.
Relations valid in this case can be found in Appendix~D.

We further consider a symmetric $ 1\times N $ coupler and
$ N $ identical detectors:
$$         
 d^{\rm noise}_{j} = d^{\rm noise},\quad
 d^{\rm dark}_j = d^{\rm dark}, \quad
 \eta _{j} = \eta ,\quad
 t_{j} = \frac{1}{\sqrt{N}}, \quad
 \mu^{\rm res}_{I,j} =  \mu^{\rm res}_I  \rightarrow
$$
\begin{equation}
d_j = d, \quad
{\cal A}_j = {\cal A} , \quad
\nu^{\rm res}_{I,j} = \nu^{\rm res}_I.
\end{equation}
The symmetric configuration provides the best results in
the exclusion of multiphoton Fock states, because ``the mean number
of photons is uniformly distributed onto all detectors''.
We also assume that postselection occurs if an arbitrary
detector beyond the $ 1\times N $ coupler detects a photon and
the rest of $ N-1 $ detectors do not register a photon.
We have in this case:
\begin{eqnarray}    
 (\hat{\rho}_{I}^{\rm mix,s})_{nn} &=& \frac{ 1-
 \nu^{\rm res}_I }{
 \exp(-\mu {\cal B}) - (1-d) \exp(-\mu {\cal A} ) }
  \nonumber \\
  & & \mbox{} \times \left\{ \exp (-\mu {\cal B})
 \exp(-\mu {\cal A} T_i) (\nu_{I}^{\rm res})^n
  f_n\left( \frac{\mu {\cal A} T_I}{\nu_{I}^{\rm res}} \right)
 \right. \nonumber \\
  & & \left. \mbox{} - (1-d) \left[ \exp(-\mu {\cal A})
 \exp(-\mu {\cal B} T_I) (\nu_I^{\rm res})^n
  f_n\left( \frac{\mu {\cal B} T_I}{\nu_I^{\rm res}} \right)
  \right] \right\} ,
  \nonumber \\
 & &
\end{eqnarray}
and
\begin{equation}  
 r_{I}^{\rm s} = N \left\{ (1-d)^{N-1}
  \exp[-\mu(1-{\cal A})]
  - (1-d)^N \exp[-\mu(1-{\cal B})] \right\} .
\end{equation}

\section{Behavior of the photon source}

The photon source is characterized by the following quantities that
are namely convenient for the description of its single-photon
character important for quantum cryptography.
A triggering probability $ p^{\rm post} $ is determined
by the probability that detection has occurred
in the signal beam:
\begin{equation}    
 p^{\rm post} = r^{\rm s}_I .
\end{equation}
A coincidence-count probability $ p^{\rm coinc} $ is given
by the conditional probability that the idler
beam contains one or more photons provided that it was triggered:
\begin{equation}    
 p^{\rm coinc}=\sum^{\infty }_{i=1} (\rho^{\rm mix,s}_I)_{ii}.
\end{equation}
A vacuum probability $ p^{\rm vac} $ determines the probability of
finding zero photons in the triggered idler state:
\begin{equation}    
 p^{\rm vac}= (\rho^{\rm mix,s}_I)_{00} .
\end{equation}
The probability of finding more than
one photon in a nonempty triggered idler state is described
by a multiphoton content $ c^{\rm multi} $:
\begin{equation}   
 c^{\rm multi}=\frac{1-\left[ (\rho^{\rm mix,s}_I)_{00}
 +(\rho^{mix,s}_I)_{11} \right] }{1- (\rho^{\rm mix,s}_I)_{00}} .
\end{equation}
Photon-number squeezing of the light is determined according
to the value of the Fano factor $ F $:
\begin{equation}    
 F = \frac{ \langle m^2 \rangle - (\langle m \rangle )^2}{
  \langle m\rangle} , \hspace{1cm}
 \langle m^2 \rangle = \sum^{\infty }_{m=1}m^{2}\,
 (\rho^{\rm mix,s}_I)_{mm} , \hspace{0.5cm}
 \langle m \rangle = \sum^{\infty}_{m=1} m \,
 (\rho^{\rm mix,s}_I)_{mm} . \nonumber
\end{equation}

The photon source operates in the ideal case as follows.
Perfect entanglement between the photons in the signal and
idler fields together with the postselection procedure
eliminates the vacuum state in the idler field. On the
other hand, a high number of ideal detectors beyond the
$ 1\times N $ coupler in the signal field suppress
the occurrence of Fock states with the photon number greater
than one in the idler field. Thus the idler field
is close to the Fock state with one photon.
Such a state is ideal for the transmission of information
in quantum cryptography. This state is also highly
nonclassical --- it exhibits photon-number squeezing.

We first consider ideal detectors ($ \eta = 1 $, $ d = 0 $)
and perfect entanglement between the signal and idler fields
($ T_S = T_I = \Theta = 1 $). A typical behavior of the
triggering probability $ p^{\rm post} $ as a function of the
mean number of pairs $ \mu $ for both one and many
detectors in the signal beam is shown in Fig.~2a.
The triggering probability $ p^{\rm post} $ grows up to unity with
incresing mean number of pairs $ \mu $ for $ N=1 $,
while it shows a maximum close to $ \mu =1 $
for large $ N $\footnote{
For an ideal photon-number-resolving measurement device,
$ p^{\rm post} $ is maximum for $ \mu = 1 $; then
$ p^{\rm post} = e^{-1} \approx 0.37 $. Comparing this value
with that in Fig.~2a for
$ N = 1000 $ detectors we get, that the $ 1 \times N $ coupler
with $ N $ detectors behaves nearly as an ideal
photon-number-resolving device.}
and then falls down to zero.
A decrease of the triggering probability $ p^{\rm post} $ for large
$ N $ is caused by the fact that fields with high intensities
have a high probability of multi-photon states which are
eliminated by the many-detector device.
The coincidence-count probability $ p^{\rm coinc} $ plotted in Fig.~2a
is unity regardless of the value of the mean number of pairs
$ \mu $ as a result of the perfect entanglement between
the signal and idler fields.
Typical experimental ranges of the mean number of pairs
$ \mu $ for cw and
pulsed-pumping regimes of the down-conversion process
(detection time 1~ns is assumed) are
also indicated in Fig.~2a.
The multi-photon
content $ c^{\rm multi} $ is shown in
Fig.~2b for several values of $ N $.
The more detectors are used in the device, the better exclusion
of multi-photon states is achieved.
The vacuum probability $ p^{\rm vac} $ is always zero in this ideal
case.
\begin{figure}    
 \begin{center}
   \epsfig{file=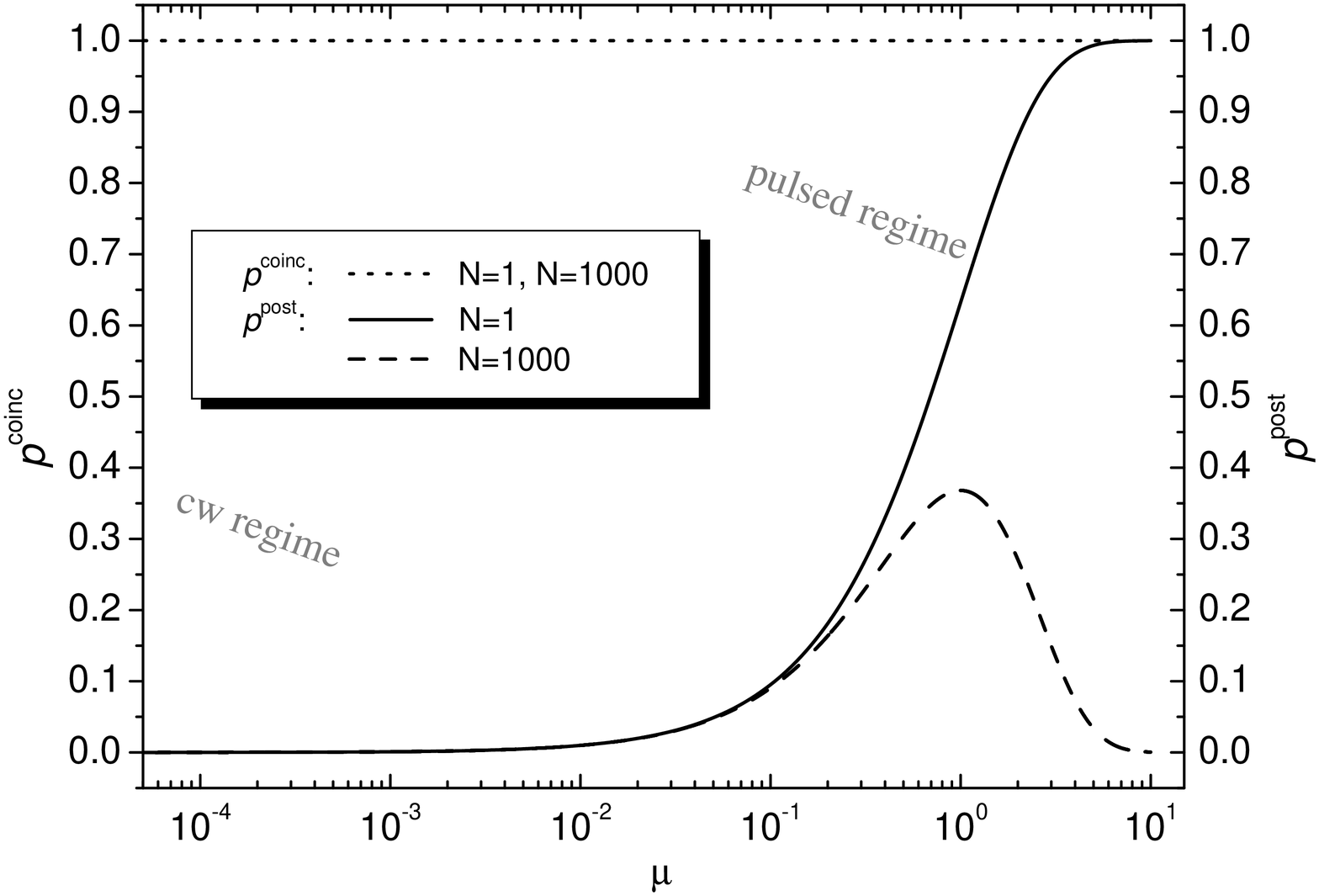,angle=0,width=0.45\hsize}
   \hspace{0.5cm}
   \epsfig{file=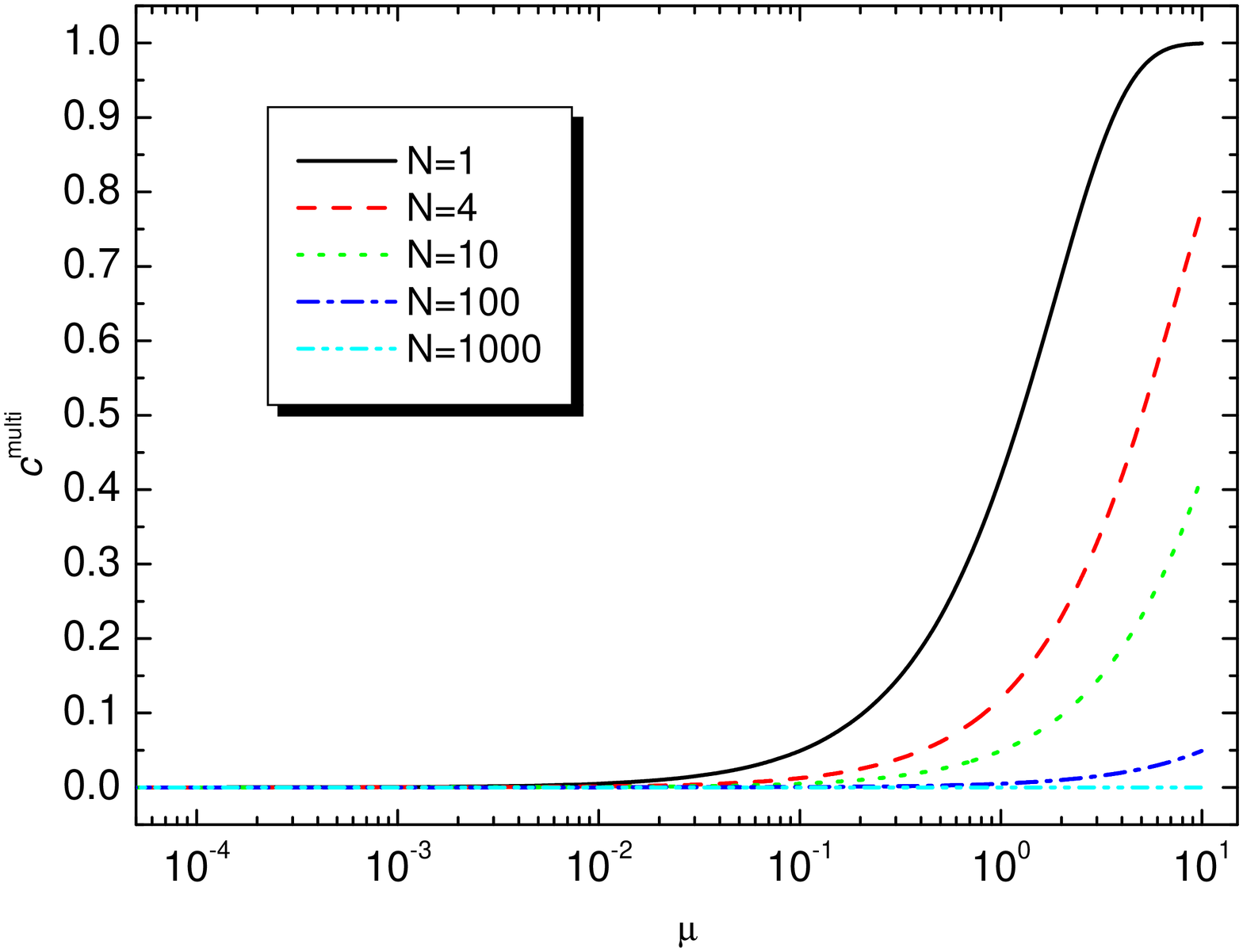,angle=0,width=0.45\hsize}
 \end{center}
 \caption{(a) Triggering probability $ p^{\rm post} $,
 coincidence-count probability
 $ p^{\rm coinc} $, and (b) multiphoton content $ c^{\rm multi} $
 for the ideal case: $ d = 0 $, $ \eta = 1 $, and
 $ T_S = T_I = \Theta = 1 $.
 Detection interval $ \tau = 1 $~ns is assumed. The curves
 showing $ p^{\rm coinc} $ for $ N=1 $ and $ N=1000 $ coincide.}
\end{figure}

We now study the influence of real detectors with non-negligible
noise and limited quantum efficiency ($ \eta < 1 $, $ d > 0 $, $
T_S = T_I = \Theta = 1 $). In general, the lower the quantum efficiency $
\eta $, the worse the exclusion of multiphoton states in the idler
field. Nonzero values of $ d $ lead in principle to the occurrence
of vacuum state in the idler field.

The triggering probability $ p^{\rm post} $ (see Fig.~3a) is now lower than
in the previous ideal case owing to losses in the postselection
device stemming from limited quantum efficiencies of the detectors.
Maximum of the triggering probability $ p^{\rm post} $ in case with many
detectors is shifted to higher values of the mean number of pairs
$ \mu $ for the same reason. The coincidence-count probability
$ p^{\rm coinc} $
approaches unity only in the high-intensity limit. It drops with
decreasing mean number of pairs $\mu$. The more the detectors,
the faster the decrease. The reason lies in the increased number of `false'
triggers in the postselection device due to dark counts of the
detectors. This fact is also reflected in the plot of the vacuum
probability $ p^{\rm vac} $ in Fig.~3b. The use of several
detectors brings only a moderate improvement in the exclusion of
multiphoton states (see Fig.~3b). The dependence of the Fano
factor $ F $ on the mean number of pairs $ \mu $ is given by the
weights of the vacuum and multiphoton contributions. The vacuum
contribution prevails for low values of $ \mu $, whereas the
multiphoton contribution is crucial for high values of $ \mu $. Photon number
squeezing with $ F<0.05 $ is achievable for the mean number of
pairs $ \mu $ in the region
$10^{-4}<\mu<10^{-1}$ and for $ N<10 $ .
\begin{figure}    
 \begin{center}
   \epsfig{file=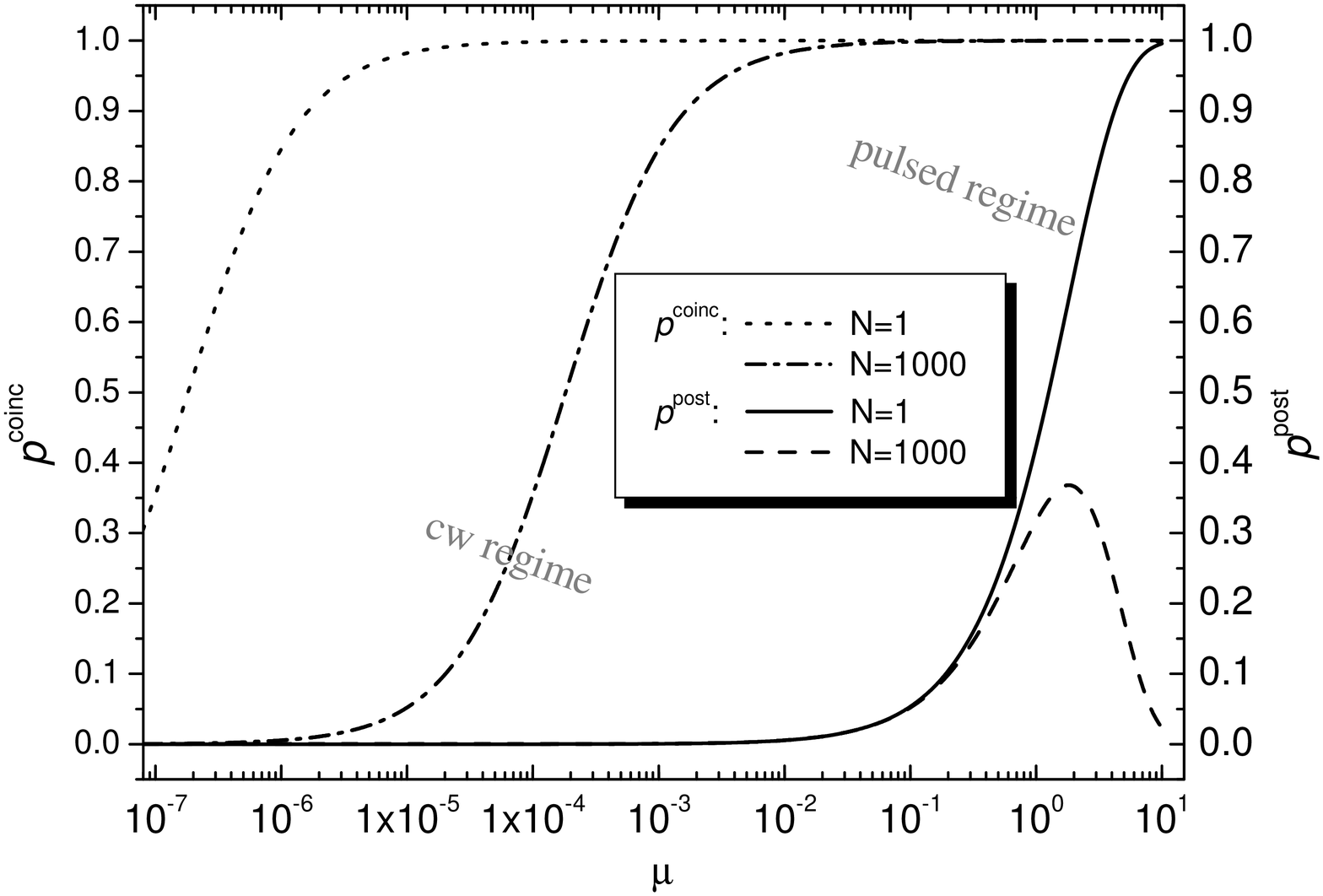,angle=0,width=0.45\hsize}
   \hspace{0.5cm}
   \epsfig{file=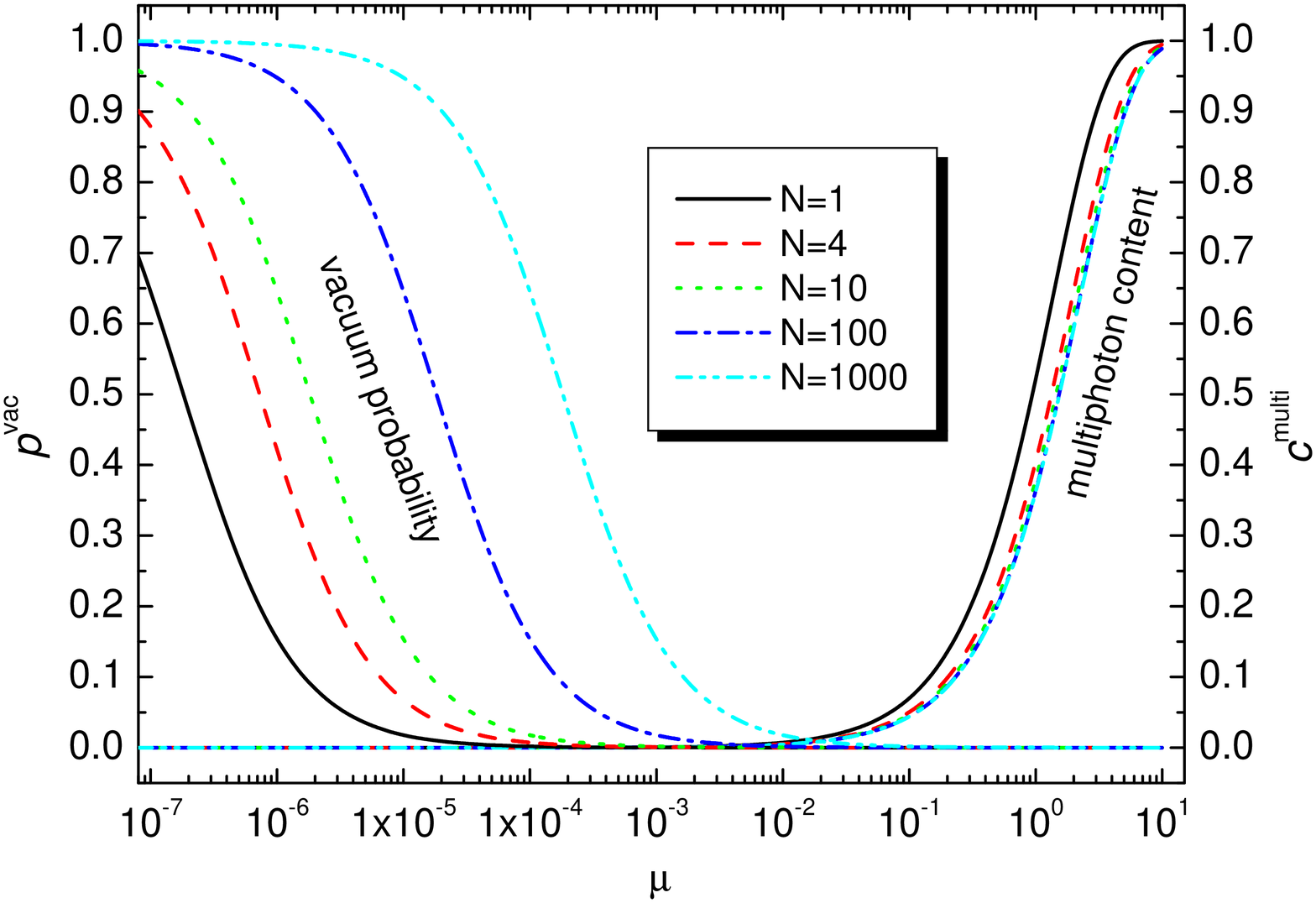,angle=0,width=0.45\hsize}
 \end{center}
 \caption{(a) Triggering probability $ p^{\rm post} $,
coincidence-count probability
 $ p^{\rm coinc} $, (b) vacuum probability $ p^{\rm vac} $, and
 multiphoton content $ c^{\rm multi} $ for the case with real detectors
 (values of parameters appropriate for silicon avalanche
 detectors are used);
 $ d^{dark}=10^{-7} $ (According to
 Eq.~(18), $ d^{\rm dark} = \mu^{\rm dark} /
 ( 1+ \mu^{\rm dark}) $, $ \mu^{\rm dark} $ being the mean
 number of dark counts. For $ \mu^{\rm dark} \ll 1 $,
 $ d^{\rm dark} \approx \mu^{\rm dark} $;
  $ \tilde{\mu}^{\rm dark} = 100 $~s$ {}^{-1} $ and the
 detection interval $ \tau = 1 $~ns are assumed),
 $ \eta =0.55 $, and $ T_{S}=T_{I}= \Theta = 1 $.}
\end{figure}

Photons in the signal and idler fields are not perfectly
entangled in a real experiment, because pairs of photons can be
broken as they propagate towards detectors
(see Appendix~C). Photons
that lost their twins then contribute to noise both in the signal
and idler beams. As a result, the coincidence-count probability
$ p^{\rm coinc} $ decreases and almost all advantages of
the many-detector device
are lost for low coupling coefficients $ T_{S} $ and  $ T_{I} $. The
dependencies of vacuum probability $ p^{\rm vac} $ and
multi-photon content $ c^{\rm multi} $ as functions of the mean
number of pairs $\mu$ for a typical experiment are plotted in
Fig.~4a. The vacuum contribution is now considerable due to
triggers by photons that lost their twins. For low $\mu$ this is
accented even more by dark counts of the detectors. The exclusion
of multiphoton states is very inefficient. The postselection
procedure works even worse in the presence of additional noise
(see Eq. (13)) caused, e.g., by misalignment of the mode-selecting
pinholes or by stray light, as documented by the dash-dot lines in
Fig.~4a. Fig.~4b shows the dependence of the Fano factor $F$ on
the mean number of pairs $\mu$
under the same conditions. Clearly, the achievable photon-number
squeezing is severely limited by the coupling coefficient
$ \Theta $, $ \Theta = T_S = T_I $
($ F $ reaches values round 0.05 for $ \Theta=1 $). The
destructive influence of the additional noise is also clearly
visible.

The crucial role of the coupling coefficient $ \Theta $ is documented
in Fig.~5 for typical values of the mean number of pairs $\mu$
in the down-conversion
experiment pumped by cw and pulsed laser. In the cw regime, good
approximations of single-photon Fock states ($F\rightarrow0$) can
be generated in the perfect coupling limit ($ \Theta \rightarrow1$),
because the vacuum state probability $ p^{\rm vac} $ can be
made very small (with low-noise
detectors and little additional noise) and the multiphoton content
$ c^{\rm multi} $ is
negligible due to low values of $\mu$. The use of several detectors in the
postselection device is not useful; the detectors even increase the total
dark-count rate. On the other hand,
the use of several detectors yields a significant
improvement for higher values of $ \Theta $ in the pulsed regime
because the exclusion of multiphoton
states from the idler field becomes efficient. It is, however, never
perfect for a realistic number of detectors. This is the reason
why not so good values of photon-number squeezing (low values of $F$) are
achievable compared to the cw regime.

\begin{figure}    
 \begin{center}
   \epsfig{file=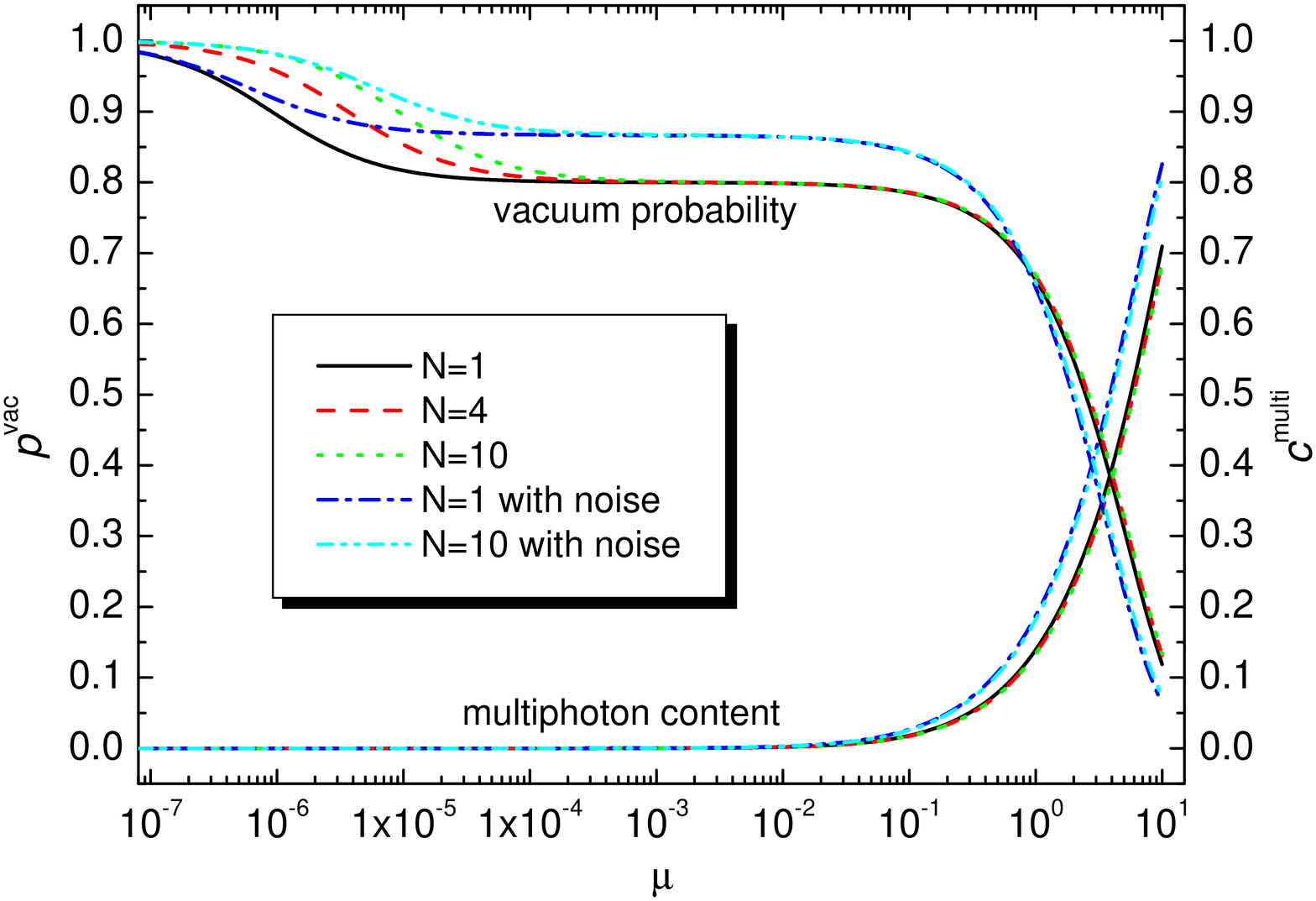,angle=0,width=0.45\hsize}
   \hspace{0.5cm}
   \epsfig{file=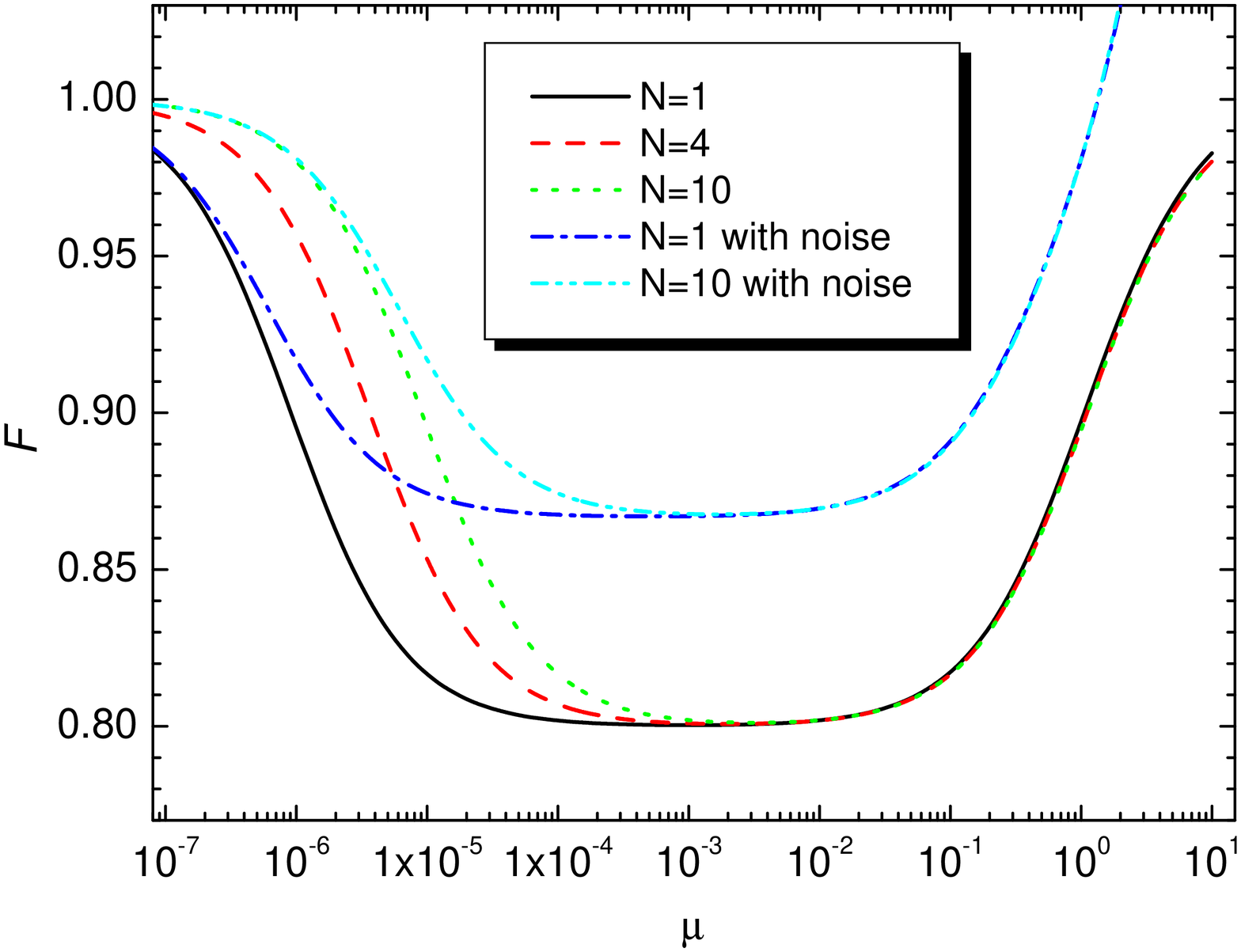,angle=0,width=0.45\hsize}
 \end{center}
 \caption{(a) Vacuum probability $ p^{\rm vac} $,
 multiphoton content $ c^{\rm multi} $, and
 (b) Fano factor $ F $ as functions of the mean number of pairs
 $\mu$; $ T_S = T_I = \Theta = 0.2 $, $ d^{\rm dark}=10^{-7} $,
 $ \eta = 0.55 $, $ \tau = 1$~ns,
 $ \mu_S^{\rm res,add} = \mu_I^{\rm res,add} = 0 $
 ($ \mu_S^{\rm res,add} = \mu_I^{\rm res,add} = 0.04 \mu $
 for curves denoted as with noise).}
\end{figure}

\begin{figure}    
 \begin{center}
   \epsfig{file=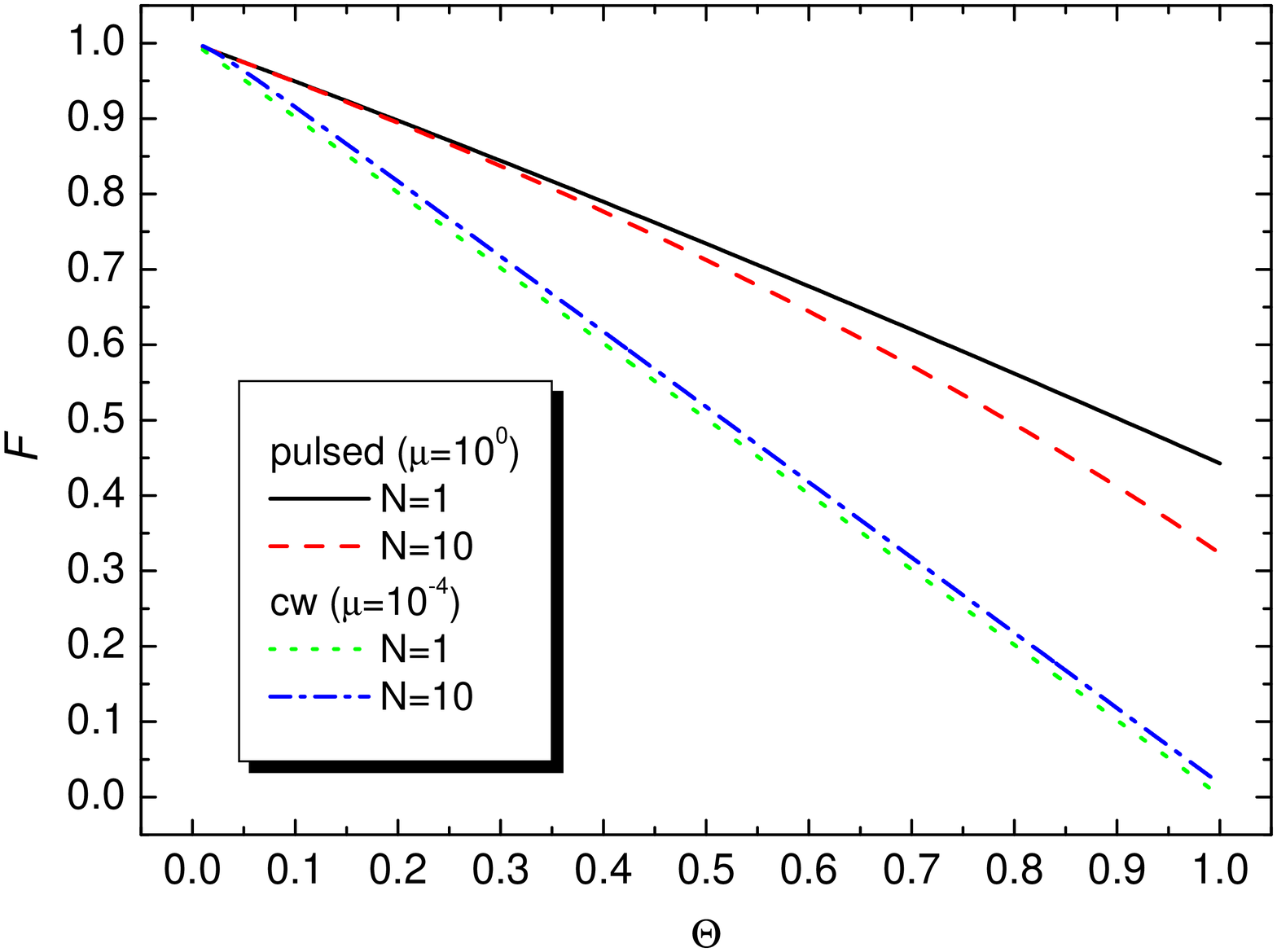,angle=0,width=0.5\hsize}
 \end{center}
 \caption{Fano factor $F$ as a function of
the coupling coefficient
$ \Theta=T_S=T_I$ for typical values of the mean number of pairs
$\mu$ in cw regime ($ \mu = 10^{-4} $)
 and in pulsed regime ($ \mu = 1 $);
 $ d^{\rm dark}=10^{-7} $, $ \eta = 0.55 $,
 $ \tau = 1$~ns, and
 $ \mu_S^{\rm res,add} = \mu_I^{\rm res,add} = 0 $.}
\end{figure}

The principle of the postselection device is well illustrated in
Figs.~6a,b where the histograms of the photon-number distribution
$ p(n,\Theta) $ as a
function of the coupling coefficient $\Theta=T_S=T_I$ assuming pulsed
pumping are
plotted. In the ideal case (Fig.~6a) employing large number of
noiseless detectors with the quantum efficiency $ \eta = 1 $,
we can see that a
perfect elimination of both multiphoton and vacuum contributions is
achieved for high values of the coupling coefficient $ \Theta $.
Using a realistic postselection
device, however, the exclusion of multiphoton contributions fails
owing to a limited number of detectors and their limited quantum
efficiencies. On the other hand an almost perfect exclusion of the vacuum
state is still achievable with today's silicon detectors.

\begin{figure}    
 \begin{center}
   \epsfig{file=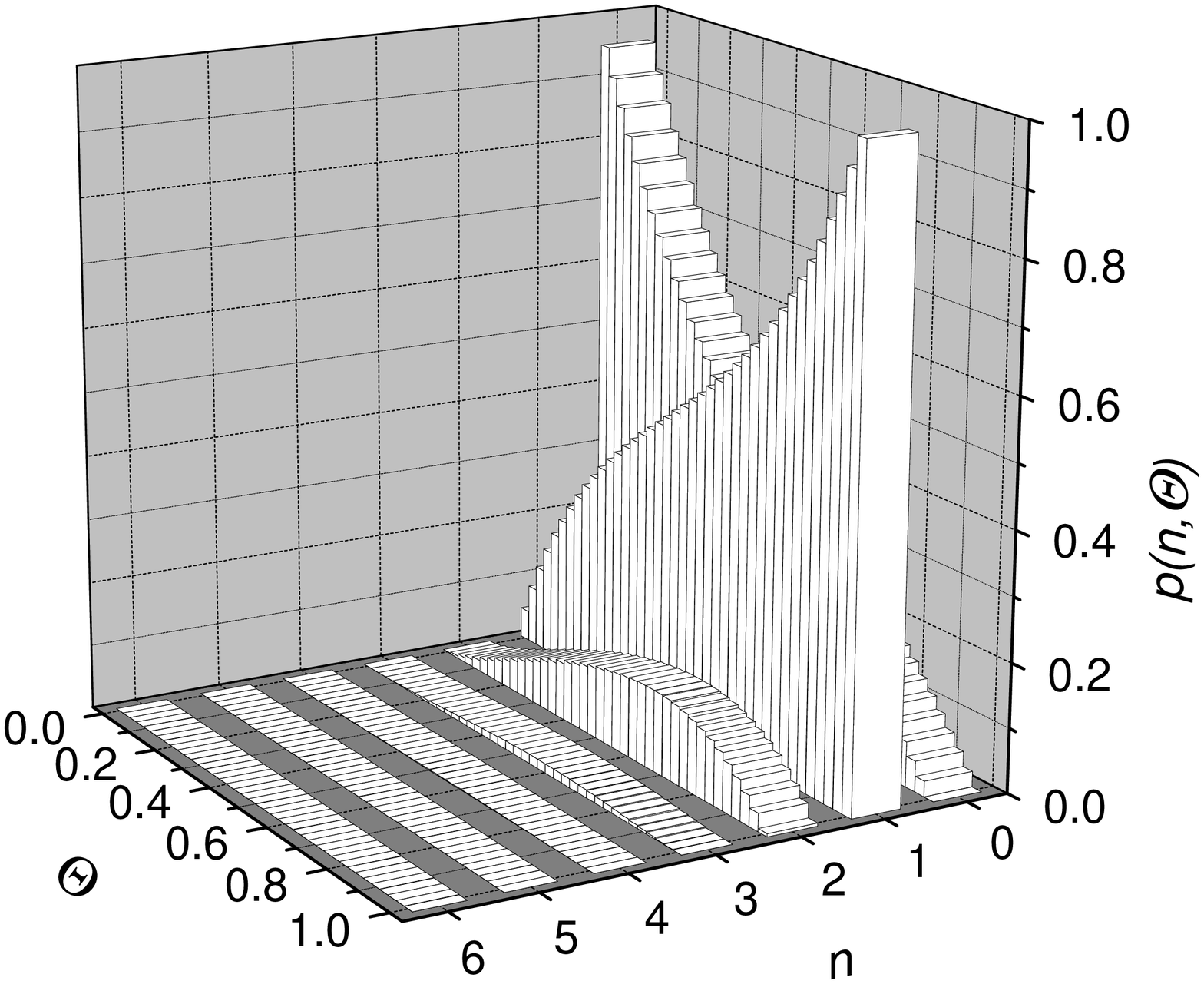,angle=0,width=0.45\hsize}
   \hspace{0.5cm}
   \epsfig{file=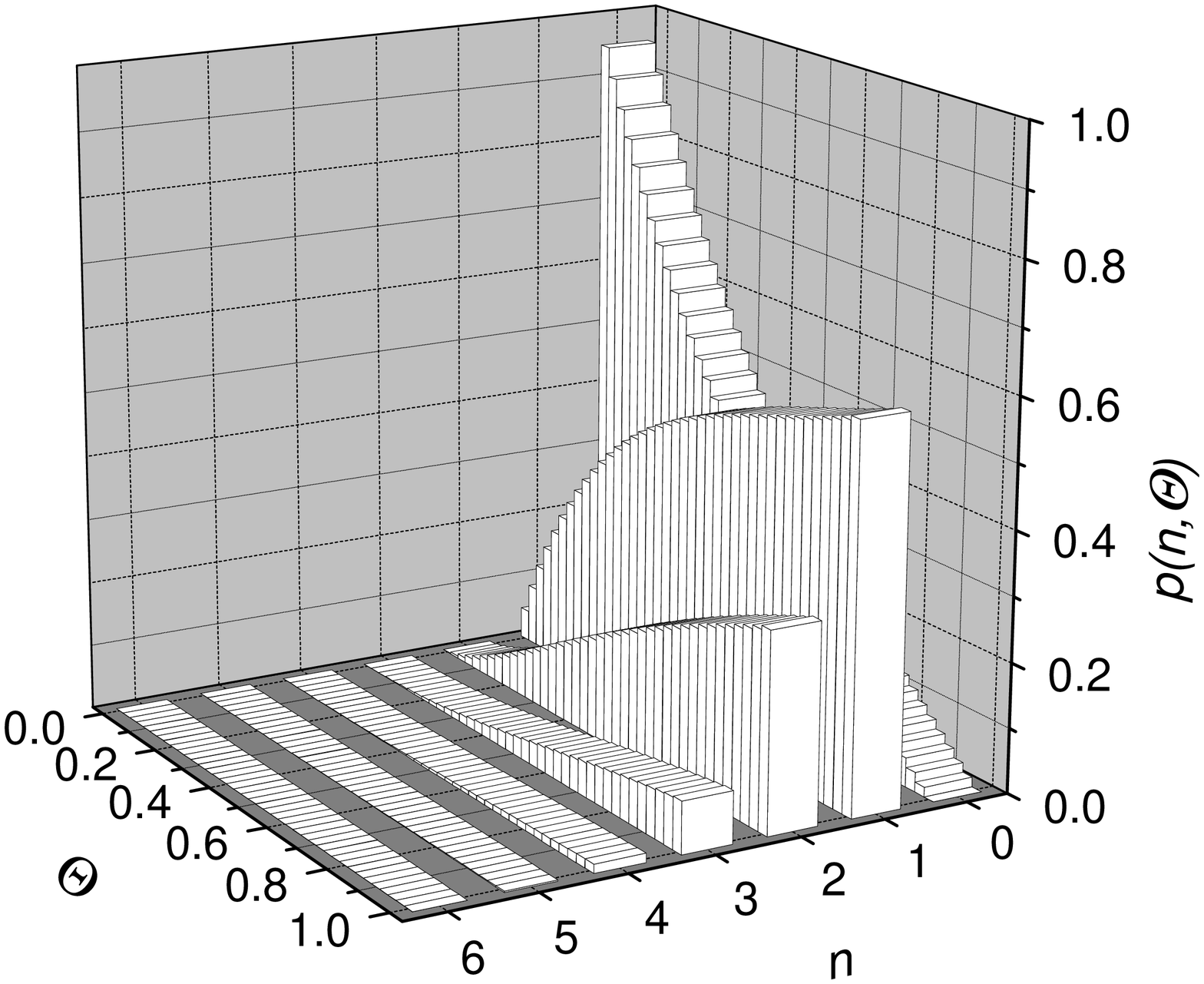,angle=0,width=0.45\hsize}
\end{center}
 \caption{Photon number distribution $ p(n,\Theta) $ ($ n $ denotes
  the photon number, $ \Theta=T_S=T_I $)
  of the state obtained in the idler beam
  if (a) $ N=100 $ ideal detectors ($d^{\rm dark}=0, \eta=1$)
  and (b) $N=10$ realistic detectors ($d^{\rm dark}=10^{-7}, \eta=0.55$)
  are used in the postselection device (symmetric $ 1 \times N $
  coupler is considered). Pulsed regime is
  considered ($\mu=1$).}
\end{figure}

\section{Use of the source for QKD}

A gap between the ultimate proofs of security of QKD and practical
systems exists, because the proofs including most general attacks
allowed by quantum mechanics \cite{mayers,lochau} still need to
make assumptions that are not implementable in the laboratory and
therefore do not yield an instruction how to build a QKD system
(one of these assumptions invoked in \cite{mayers} is the existence of a
single-photon source). However, if we slightly weaken our security
requirements and limit the eavesdropper to attacks on single
particles only (omitting the so-called coherent attacks), there is
a proof due to L\"utkenhaus that is in correspondence with current
experimental techniques \cite{lutkenhaus}.
Here the eavesdropper is allowed to use any general quantum-mechanical
measurement on single qubits (or ancillas bound to single qubits)
including identification and efficient splitting of multiphoton states
together with the possibility to store
the states until measurement bases are announced in the public discussion.

According to this proof a QKD system may expediently be
characterized by the quantity called gain $ G $ \cite{lutkenhaus}.
Gain $ G $ characterizes the fraction of a bit of the key
established by the QKD procedure per qubit sent over a quantum
channel. Gain $ G $ is determined as follows:
\begin{equation}       
\label{gain}
G=\frac{1}{2}\, p^{\rm post}\, p^{\rm exp}\, \left( 1-c^{\rm EC}-
c^{\rm PA}\right) .
\end{equation}
Here $ p^{\rm post} $ is the postselection probability of the source
($ p^{\rm post} = r^{\rm s}_I $ given in Eq.~(22) in our model),
$ p^{\rm exp} $ is the probability of detection at the receiving station of
QKD (usually called Bob) and $ c^{\rm EC} $ and $ c^{\rm PA} $
are error correction and privacy
amplification terms, respectively
(for details, see \cite{lutkenhaus}).
Gain $ G $ is closely connected to key generation rate:
the higher the gain $ G $, the higher the key generation rate.

Taking into account only single-particle attacks,
the $ c^{\rm PA} $ term can be expressed as \cite{fuchs,lut2}:
\begin{equation}           
\label{PA}
c^{\rm PA}=1-\frac{p^{\rm exp}-p^{\rm multi}}{p^{\rm exp}}\left\{
1-\log _{2}\left[ 1+4\, e\frac{p^{\rm exp}}{p^{\rm exp}-p^{\rm multi}}-
4\left( e\frac{p^{\rm exp}}{p^{\rm exp}-p^{\rm multi}}\right) ^{2}\right]
 \right\} ,
\end{equation}
where $ e $ is the bit error rate measured at Bob's station,
$$
p^{\rm multi}=1-\sum_{n=0}^\infty \left[ \left( 1-T_{\rm ALICE} \right)^n
+ n T_{\rm ALICE} \left( 1-T_{\rm ALICE} \right)^{n-1} \right]
(\rho _{I}^{mix,s})_{nn}
$$
denotes the probability of multiphoton
states at the beginning of the transmission line (after passing
through Alice's device with the transmission coefficient
$T_{\rm ALICE}$) and $
p^{\rm exp}=p^{\rm exp}_{\rm s}+ d_{\rm BOB}^{\rm dark}-p^{\rm
exp}_{\rm s} d_{\rm BOB}^{\rm dark} $ stands for the expected rate of
Bob's detections. In the latter relation $ d_{\rm BOB}^{\rm dark}
$ is dark-count rate of Bob's detector and $ p^{\rm exp}_{\rm s} $
means signal rate at Bob's station. The signal rate
$ p^{\rm exp}_{\rm s} $ can be expressed as:
\begin{equation}          
\label{psig}
p_{\rm s}^{\rm exp}=\sum _{j=1}^{\infty }
(\rho^{\rm mix,s}_{I})_{jj} \sum _{l=1}^{j}\left( \begin{array}{c}
j\\
l
\end{array}\right) \left( T_{TL}T_{\rm ALICE}\, \eta _{\rm BOB}\right) ^{l}
\left( 1-T_{TL}T_{\rm ALICE}\, \eta _{\rm BOB}\right) ^{j-l}.
\end{equation}
The symbol $ T_{TL}=10^{\left( -\alpha L+l_{\rm BOB}\right) /10} $
denotes the transmission coefficient of the transmission line\footnote{We
consider optical fiber serving as the quantum channel. Note that
attempts are being made to build a free space QKD \cite{hughes}.},
$ \alpha $ is fiber-attenuation factor, $ L $ means fiber length,
$ l_{\rm BOB} $ denotes losses of Bob's apparatus, and $ \eta _{\rm
BOB} $ stands for quantum efficiency of Bob's detector. The bit
error rate $ e $ in the absence of an eavesdropper is caused
either by physical imperfections (at a rate $ c $) or by dark
counts of Bob's detector at the rate 0.5 and we therefore have:
\begin{equation}             
\label{e}
e=\frac{c\, p_{\rm s}^{\rm exp}+\frac{1}{2}
d^{\rm dark}_{\rm BOB}-\frac{1}{2}c\, p_{\rm s}^{\rm exp}
d^{\rm dark}_{\rm BOB}}{p^{\rm exp}}.
\end{equation}
The error correction term $ c^{\rm EC} $ is expressed by the
formula \cite{brassal}:
\begin{equation}            
\label{EC}
c^{\rm EC} \approx -1.16\, \left[ e\log _{2}e+\left( 1-e\right)
\log _{2}\left( 1-e\right) \right]
\end{equation}
valid for $ e\leq 0.05 $.

Using dim coherent states there is always an optimum value of the
source mean photon number $ \mu $ (see Fig.~7).
\begin{figure}    
 \begin{center}
   \epsfig{file=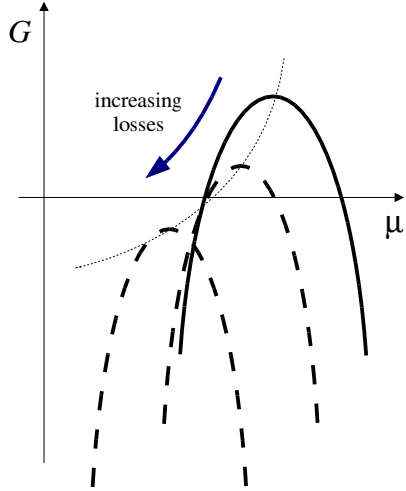,angle=0,width=0.3\hsize}
 \end{center}
 \caption{A typical curve (solid) characterizing
 the dependence of the gain $ G $ on the mean number of photons
 $ \mu $ of a coherent Poissonian source. Secure QKD is possible
 if $ G \ge 0 $. Dashed curves show how the gain $ G $ behaves when
 losses on the
 transmission line (or length of the fiber) increase. The lower
 the gain $ G $, the lower the optimum mean photon number of the source.
 If the losses are too high, secure QKD is impossible.}
\end{figure}
There is a high vacuum content in the signal quantum states for low
values of the source mean photon number $ \mu $ and so the quantity
$ c^{\rm EC} $ rapidly grows because physical
noises of the detector on Bob's side become stronger than the signal
itself. Each noise count contributes by 50\% error rate. On the other hand,
the contribution of multiphoton states in the signal field for high values
of the source mean photon number $ \mu $
requires high values of
$ c^{\rm PA} $ term which again make the gain $ G $ negative at some point.
If the gain $ G $ is positive for the optimum source mean photon
number $ \mu $, secure QKD can be performed.
There is a maximum allowed amount of losses (or a maximum
achievable length of the fiber) for which the required security
is still preserved (though at a very low gain).
Unfortunately the distances achievable with dim coherent states are rather
low, about 8~km using the 800~nm communication window in optical fibers, or
about 25~km in the 1550~nm region \cite{lutkenhaus}.
Both communication windows have their caveats.
While silicon detectors currently used at 800~nm exhibit
very low noise (dark-count rates less than 100 s$ ^{-1} $) and
high quantum efficiencies ($ \eta >0.5 $),
the losses of the transmission line are very high
($ \sim 2.5 $~dB~km$ ^{-1} $).
Just the opposite is available at 1550~nm: transmission-line losses below
0.2~dB~km$ ^{-1} $ and detectors with quantum efficiencies below
0.2 and with $ 10^{4} \div 10^{5} $ dark counts per second.

L\"utkenhaus considered an idealized model of the source
based on postselection from entangled photon pairs and found out
\cite{lutkenhaus} that the communication
distance might extend up to 110~km when a local postselection device operating
at 800 nm is used and the transmission line is in the low-loss
1550~nm window. This represents an optimum choice with current technology.

We first analyze an idealized postselection device in our model.
We thus consider noiseless detectors ($ d^{\rm dark}=0 $) with
unity quantum efficiency ($ \eta =1 $) and perfect coupling ($
T_{S}=T_{I}= \Theta = 1 $). The idler beam is led from Alice ($T_{\rm
ALICE}=0.79$) to Bob's realistic detector
($ \eta _{\rm BOB}=0.18 $, $ d_{\rm BOB}^{\rm dark}=2\times10^{-5}
$) using 1550~nm transmission line ($ \alpha
=0.2 $~dB~km$ {}^{-1} $, $ c=0.01 $). We can see in Fig.~8 that the upper limit of the
communication distance extends up to 161~km. This is more than six
times the distance achievable with coherent states. This is mainly
because the transmitted quantum states contain only a small
contribution of the vacuum state. The use of more detectors in the
postselection device (this improves the photon-number resolution)
does not lead to any further extension of the communication
distance. The reason is that the maximum distance is given by the
signal-to-noise ratio at Bob's detector that becomes too low when
$ p^{\rm exp}_{\rm s} $ drops to the order of $ d_{\rm BOB}^{\rm
dark} $, i.e. deep down below unity. The multiphoton content
$ c^{\rm multi} $ in
the signal field is negligible in this case. Nevertheless, the
improvement of the photon-number resolution leads to an
improvement of the gain $ G^{\rm opt} $ up to several
orders of magnitude (see Fig.~8) and therefore to an improvement
of the key generation rate.
\begin{figure}    
 \begin{center}
   \epsfig{file=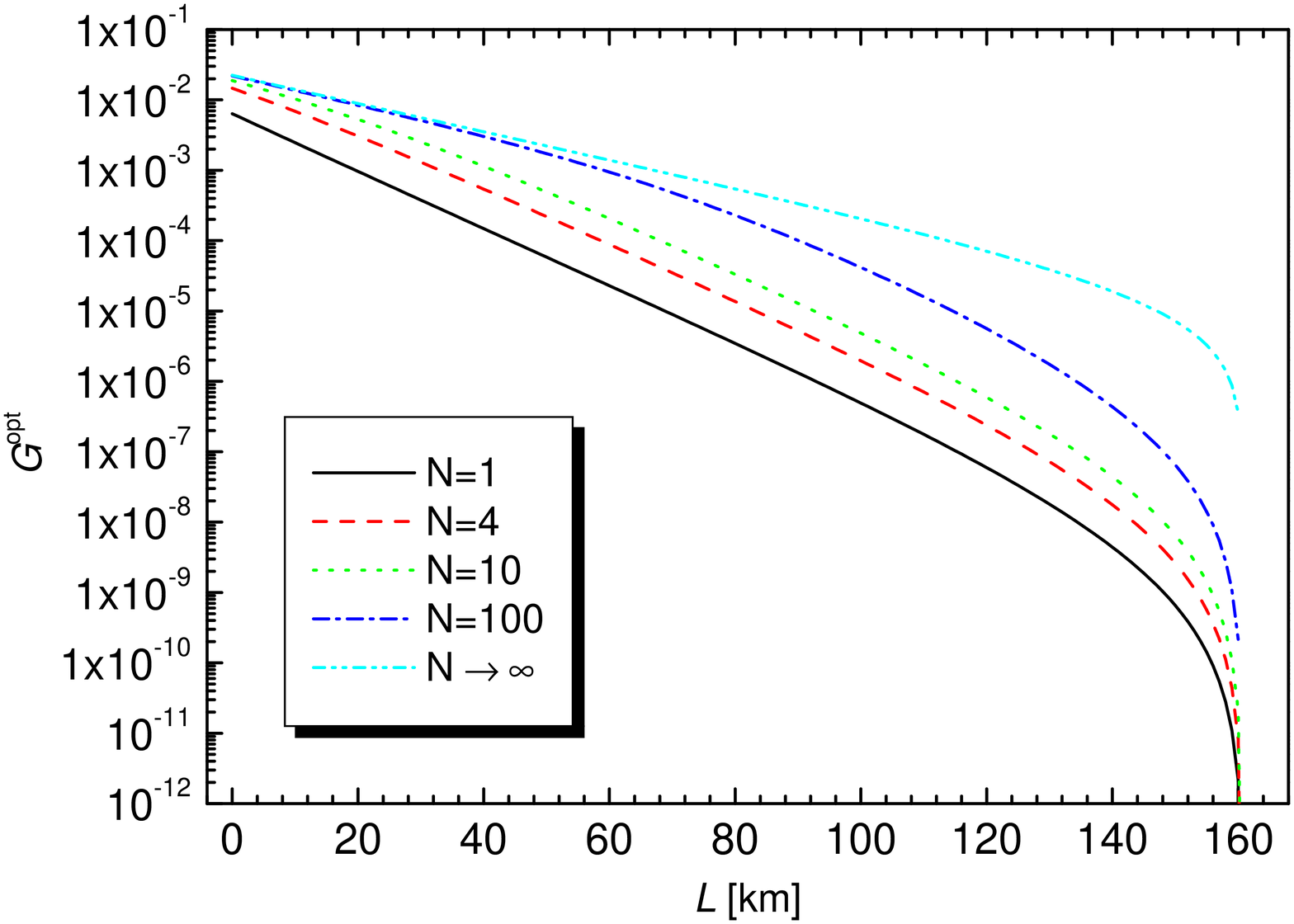,angle=0,width=0.55\hsize}
 \end{center}
 \caption{Optimum gain $G^{\rm opt}$ of the
down-conversion source as a function of the transmission distance
$ L $
using ideal postselection device ($ \eta =1 $, $ d^{\rm
dark}=0 $, $ T_{S}=T_{I}=\Theta=1 $) and realistic transmission line at
1550~nm ($T_{\rm ALICE}=0.79$, $ \alpha =0.2 $~dB~km$ {}^{-1} $, $
c=0.01 $, $ \eta _{\rm BOB}=0.18 $, $ d_{\rm BOB}^{\rm
dark}=2\times10^{-5} $); $ \mu^{\rm res, add}_{S} =
\mu^{\rm res, add}_{I} = 0 $. The maximum achievable transmission
distance is characterized by the drop of the optimum gain $G^{\rm opt}$.}
\end{figure}

The dependence of the optimum mean number of pairs
$ \mu^{\rm opt} $ as a function of the
transmission distance $ L $ changes significantly if parameters
appropriate for a realistic postselection device are considered.
If the coupling coefficient $ \Theta $ of the photon pairs is set to 0.2
($ \Theta = T_{S} = T_I = 0.2 $, this value is typical for current down-conversion
experiments, cf.~Appendix~C) and parameters of realistic detectors
are used ($\eta=0.55$, $d=10^{-7}$), the maximum communication
distance drops down to about 120~km (see Fig.~9a). This
distance is still significantly better than that achieved with
coherent states. The use of more detectors in the postselection
device brings, however, no advantage (curves for $N=1,4,10$ almost
coincide in Fig. 9a). A closer view (see the inset in Fig.~9a) shows that
it causes even a slight drop in the
optimum key generation rate for short distances.
The reason lies in
the fact that the efficiency of the exclusion of multiphoton
states is very low due to low values of the
coupling coefficient $ \Theta $ and the negative
influence of detector noise appears to be more significant.
\begin{figure}    
 \begin{center}
   \epsfig{file=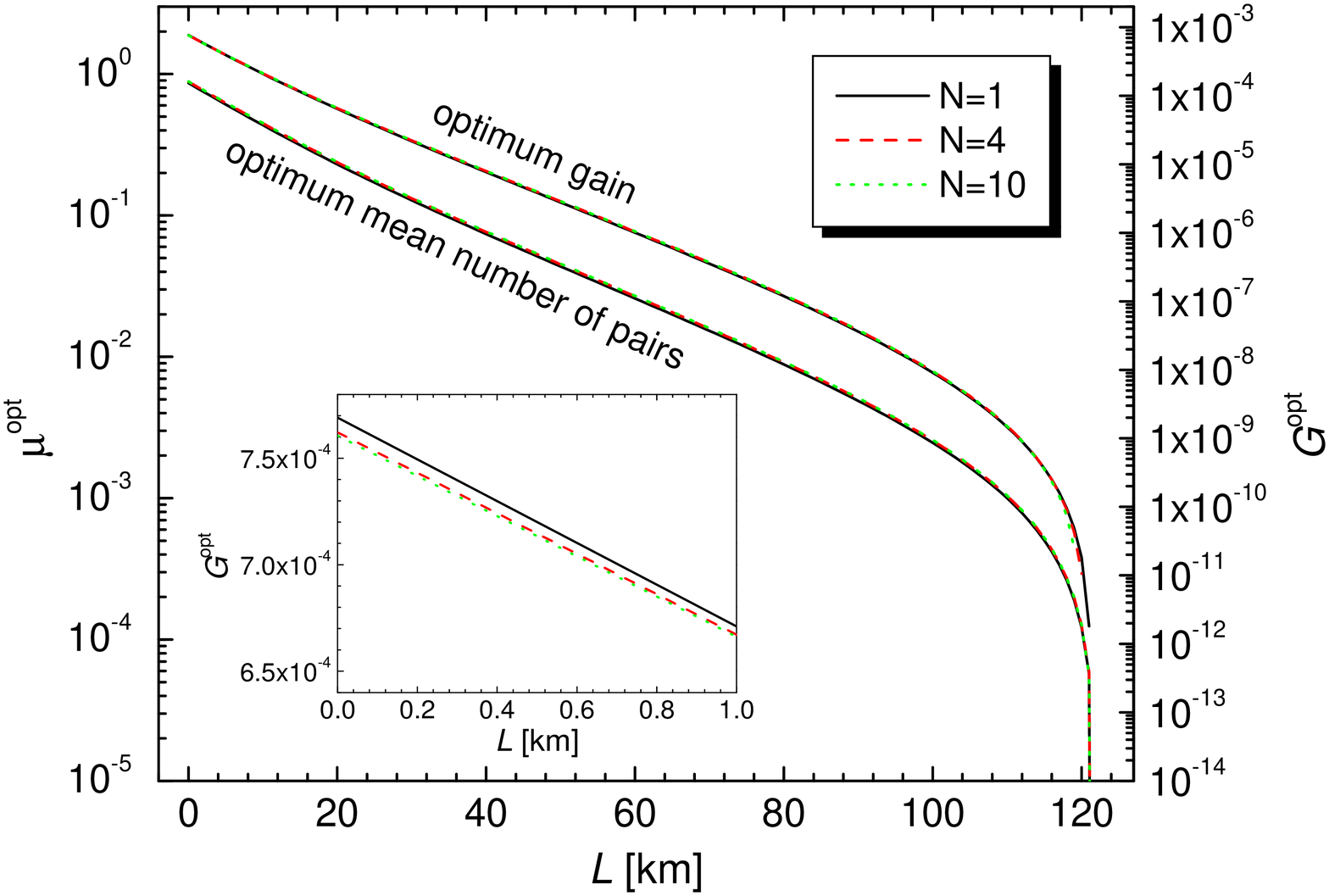,angle=0,width=0.45\hsize}
   \hspace{0.5cm}
   \epsfig{file=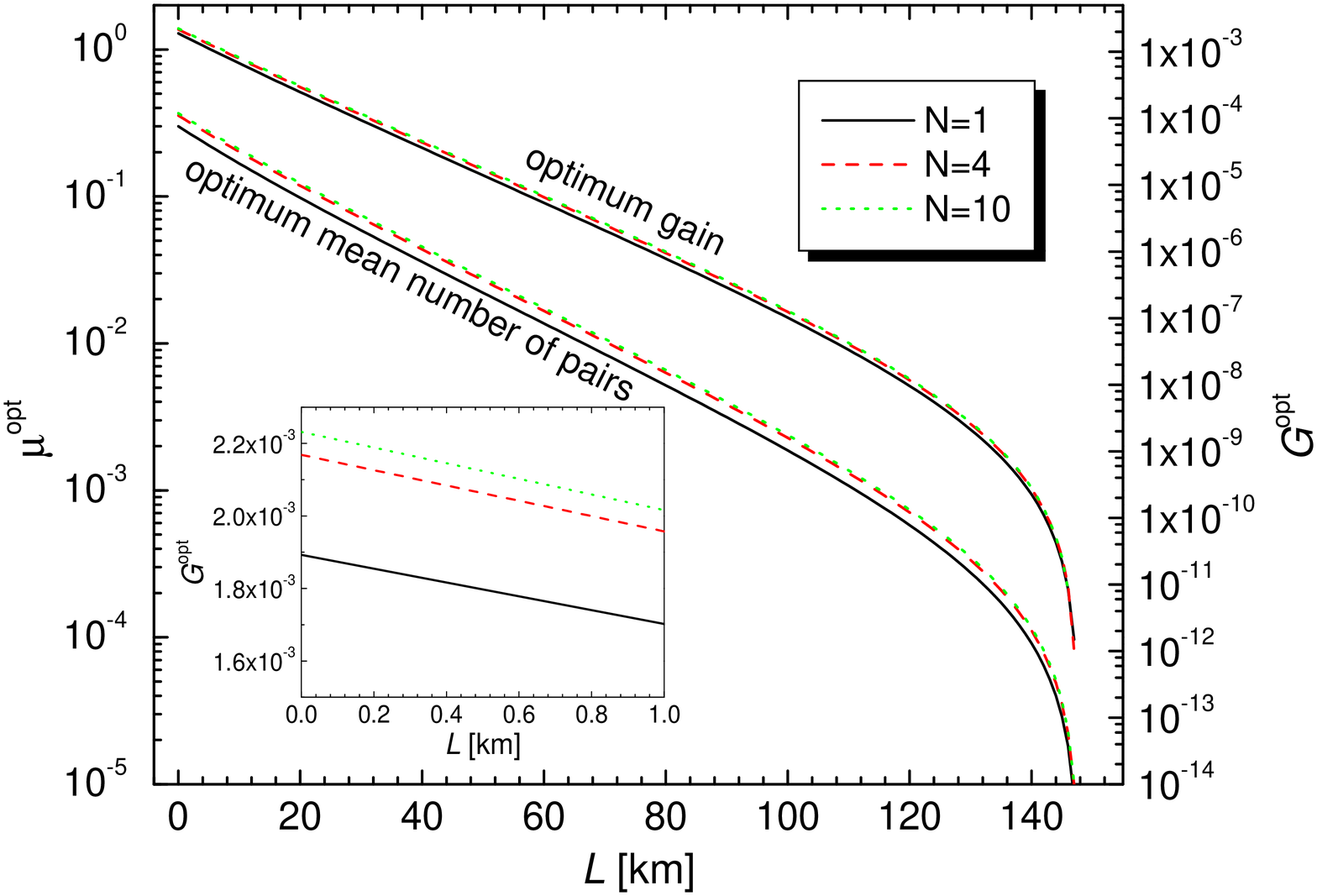,angle=0,width=0.45\hsize}
 \end{center}
 \caption{Optimum mean number of pairs $\mu^{\rm opt}$
of the down-conversion source and optimum QKD gain $G^{\rm opt}$
in the dependence of the transmission distance $ L $
(a) for realistic postselection device at 800~nm
($ \eta =0.55 $, $ d^{\rm dark}=10^{-7} $, $ T_{S}=T_{I}=\Theta=0.2 $) and
realistic transmission line at 1550~nm ($T_{\rm ALICE}=0.79$,
$ \alpha =0.2 $~dB~km$ {}^{-1} $, $ c=0.01 $, $ \eta _{\rm BOB}=0.18 $,
$ d_{\rm BOB}^{\rm dark}=2\times10^{-5} $) and
(b) for improved
postselection device using best current experimental skills
($ \eta =0.7 $, $ d^{\rm dark}=2\times10^{-8} $,
 $ T_{S}=T_{I}=\Theta=0.6 $)
and realistic transmission line at 1550~nm ($T_{\rm ALICE}=0.79$,
$ \alpha =0.2 $~dB~km$ {}^{-1} $, $ c=0.01 $, $ \eta _{\rm
BOB}=0.18 $, $ d_{\rm BOB}^{\rm dark}=2\times10^{-5} $);
$ \mu^{\rm res, add}_{S} = \mu^{\rm res, add}_{I} = 0 $.
Curves for $ N=1,4,10 $ in Fig. 9a) almost
coincide. $ G^{\rm opt} $ for short distances $ L $ are shown in
insets.}
\end{figure}

Increase in the key generation rate is
achievable with today's best technology (see Fig.~9b). Using
parameters of a recent down-conversion experiment
\cite{weinfurter} where a significant improvement of the coupling
coefficient $ \Theta $ has been achieved, we find the maximum achievable
transmission distance to be about 148 km. Moreover, the optimum key
generation rate can now be increased employing $N>1$ detectors in
the postselection device and is about three times higher than
that for the values of parameters used in Fig.~9a.

A crucial role of the coupling coefficient $ \Theta $ of photon pairs
is illustrated in Fig.~10.
\begin{figure}    
 \begin{center}
   \epsfig{file=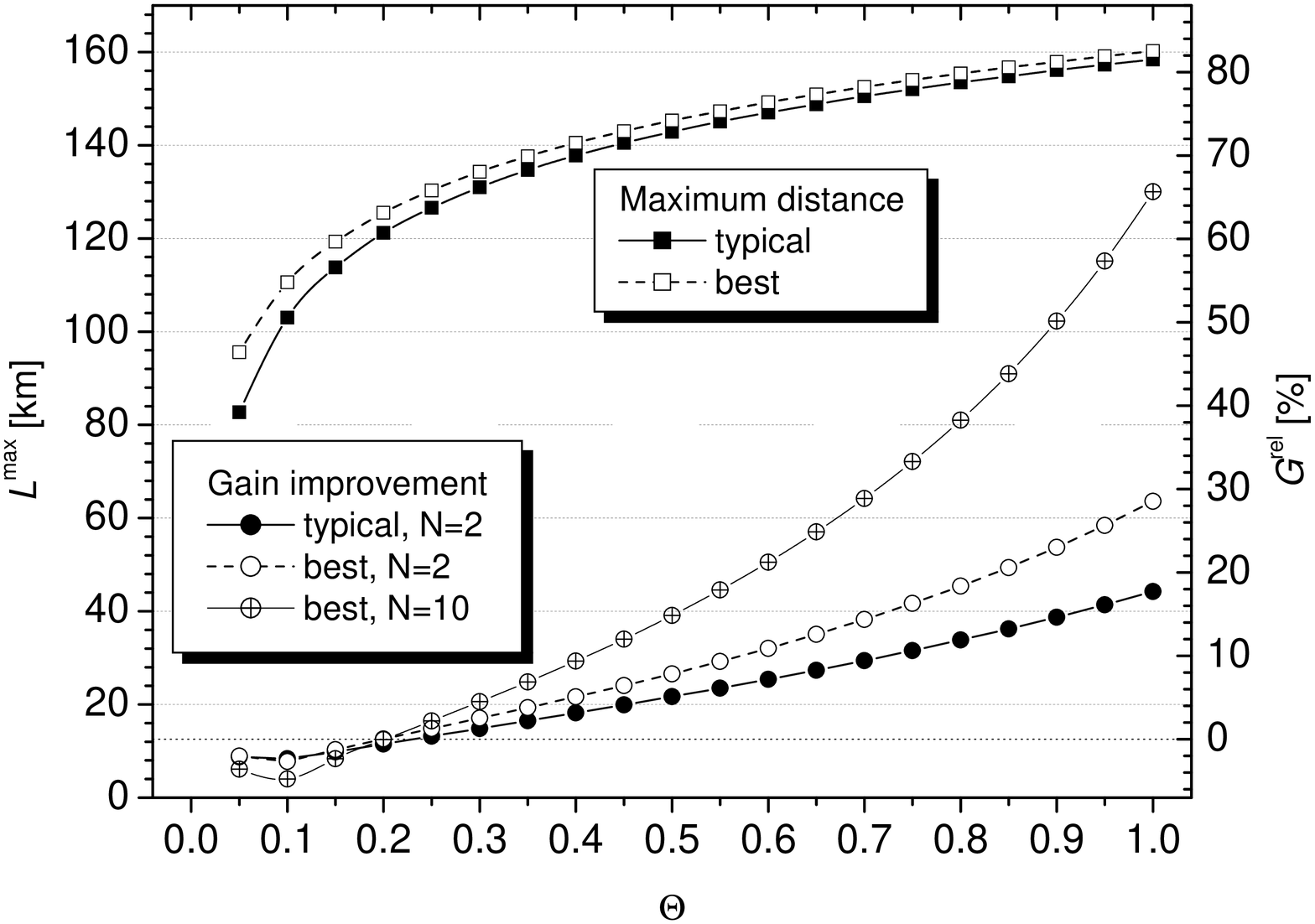,angle=0,width=0.55\hsize}
 \end{center}
 \caption{Maximum achievable transmission
distance $ L_{\rm max} $ (rectangles) as a function of the coupling
coefficient $ \Theta $ ($ \Theta = T_S = T_I $)
for realistic postselection device at 800~nm using typical values
for current down-conversion experiments (filled symbols) ($ \eta
=0.55 $, $ d^{\rm dark}=10^{-7} $, $\mu_S^{\rm res,
add}=\mu_I^{\rm res, add}=0.04 \mu$, $N=1$) and best today
achievable values of parameters (open symbols) ($ \eta =0.7 $, $ d^{\rm
dark}=2\times10^{-8} $, $\mu_S^{\rm res, add}=\mu_I^{\rm res,
add} = 0$, $N=1$). In both cases, a realistic transmission line
at 1550~nm is employed ($T_{\rm ALICE}=0.79$, $ \alpha =0.2
$~dB~km$ {}^{-1} $, $ c=0.01 $, $ \eta_{\rm BOB}=0.18 $, $ d_{\rm
BOB}^{\rm dark}=2\times10^{-5} $). The curves with circles show
the relative gain improvement $ G^{\rm rel} $
reached with the postselection device
with $ N=2 $ detectors and values of parameters typical for current
down-conversion experiments (filled circles),
with $ N = 2 $ detectors and best today achievable values of parameters
(open circles), and with $ N = 10 $ detectors and best today achievable
values of parameters (crossed circles) compared to the case with $ N=1 $
at short communication distances ($L=0$).}
\end{figure}
If values of
$ \Theta $ are greater than certain
value $ \Theta_{\rm min} $ (for our parameters
$ \Theta_{\rm min} \approx 0.2 $) then
the better the
coupling, the longer the communication distance.
Moreover, the better the coupling, the more efficient the
resolution of multiphoton states. This then results in
the gain improvement provided that more detectors are used.
To be more specific, the use of two detectors in a QKD system
characterized by typical values of parameters (see Fig.~10) results
in the improvement of the relative gain $ G^{\rm rel} $
by more than 10 \% provided that the coupling coefficient
$ \Theta $ is greater than $ 0.7 $. Values of the
coupling coefficient $ \Theta $ have to be greater
than $ 0.55 $ in a QKD system characterized
by best today available values of parameters (see Fig.~10). The use of more
than two detectors in this case results in greater values of relative
gain $ G^{\rm rel} $, as is documented in Fig.~10 for $ N=10 $.

We note that the above mentioned expression for the
gain $ G $ in Eq.~(28) can be used for the determination of
an optimum combination
of elements with given characteristics (transmission coefficents,
quantum efficiencies of detectors, noises) in a practical
implementation of a quantum key distribution system.

\section{Conclusions}

We have suggested a source of single-photon states using
a source of entangled photon pairs (nonlinear crystal with
parametric down-conversion) and postselection
in one of the entangled beams.
Based on an approximate photon-number measurement performed
with a $ 1\times N $ coupler and $ N $ detectors in the signal
beam, some realizations of the state in the idler field are selected.

An ideal device (perfect alignment of the setup, noiseless
detectors with quantum efficiency one) provides nearly
single-photon states. Real devices generate states with worse
statistics. Dark counts of the detectors and noise coming from
decorrelated photons and stray light increase the weight of
the vacuum state in the postselected (idler) field. Limited
quantum efficiencies of the detectors as well as noise
prevent from a perfect exclusion of multiphoton states
in the postselected field. However, a field close to
a single-photon state may be generated assuming
good coupling of photons in the setup and pulsed pumping
of the down-conversion process; low-noise detectors have to be
used for cw pumping. Such a field is considerably squeezed
in photon number (it has sub-Poissonian statistics) and
provides a useful source for quantum cryptography.

The suggested source with ideal values of parameters extends the maximum
communication distance of QKD about six times
(compared to a traditional coherent Poissonian source) up to 160~km.
The source with currently achievable values of parameters may be used
for the communication distances up to 120~km.
Using the best values of parameters available today, the maximum
communication distance extends up to 150~km.
The maximum communication distance is practically the same
for different numbers of detectors in the signal beam. However,
the higher the number of detectors in the signal beam, the higher
the gain and therefore also the transmission rate.
The quality of coupling of the entangled photon pairs is a crucial parameter
both
for achieving long communication distances
and optimum transmission rates.
Improvement of the coupling is a challenge for experimentalists.

\section{Acknowledgement}

This work was supported by the Ministry of Education of the Czech
Republic (projects no.~LN00A015 and no.~19982003012).

\appendix

\section{Detection operator including effects of noise}

We assume that the signal field with the density matrix
$ \hat{\rho}_S $
is mixed at the detector with a statistically independent
noisy field with the density matrix $ \hat{\rho}_R $. Detection
operator $ \hat{P}_{SR} $ describing detection of a photon either
from the signal or the noisy field has the form:
\begin{eqnarray}      
 \hat{P}_{SR} &=& \sum_{n=1}^{\infty} \left[ 1 - (1-\eta)^n \right]
  |n\rangle_S {}_S\langle n| \hat{1}_R  \nonumber \\
  & & \mbox{} + \sum_{n=0}^{\infty} (1-\eta)^n |n\rangle_S {}_S\langle n|
  \sum_{k=1}^{\infty} \left[ 1 - (1-\eta)^k \right]
  |k\rangle_R {}_R\langle k| ,
\end{eqnarray}
where $ |n\rangle_S $ ($ |k\rangle_R $) denotes a Fock state of
the signal (noisy) field and $ \eta $ stands for the quantum
efficiency of the detector.

Detection operator $ \hat{P}_S $ relevant to the signal field
is obtained from the detection operator $ \hat{P}_{SR} $ in
Eq.~(A1) by tracing over the noisy-field space:
\begin{eqnarray}      
 \hat{P}_{S} &=& {\rm Tr}_R \left\{ \hat{P}_{SR} \right\}
   \nonumber \\
      &=& \sum_{n=1}^{\infty} \left[ 1 - (1-\eta)^n \right]
  |n\rangle_S {}_S\langle n| + d \sum_{n=0}^{\infty} (1-\eta)^n
  |n\rangle_S {}_S\langle n| .
\end{eqnarray}
The constant $ d $ has the form:
\begin{equation}    
 d = \sum_{k=1}^{\infty} \left[ 1 - (1-\eta)^k \right]
 {}_{R} \langle k| \hat{\rho}_R |k\rangle_R .
\end{equation}
The relation $ 0 \le d \le 1 $ follows from Eq.~(A3).
The higher the mean number of photons in the noisy field,
the higher the value of $ d $.

If several noise sources are present at the detector, the
constant $ d $ in Eq.~(A2) is defined as follows:
\begin{equation}   
 d = d_{R1} + (1-d_{R1})
 d_{R2} + (1-d_{R1}) (1-d_{R2})
 d_{R3} + \ldots ,
\end{equation}
where the constants $ d_{R1} $, $ d_{R2} $,
$ d_{R3} $, $ \ldots $ describe
the influence of noisy fields $ R_1 $, $ R_2 $, $ R_3 $,
$ \ldots $
and are determined according to Eq.~(A3).

\section{Statistics in multimode parametric frequency down-conversion}

Expanding the interacting fields into harmonic plane waves,
the interaction Hamiltonian $ \hat{H}_{\rm int} $ of the process
of spontaneous parametric frequency down-conversion can be written
in the form \cite{mandelwolf,disper,perinajr}:
\begin{eqnarray}     
 \hat{H}_{\rm int}(t) &=& C_{\rm int}
 \int_{-L}^{0} dz \, \sum_{k_p} \sum_{k_s}
 \sum_{k_i} \chi^{(2)}
 {\cal E}_p^{(+)}(0,\omega_{k_p}-\omega^0_p)
 \hat{a}_s^{\dagger}(k_s)
 \hat{a}_i^{\dagger}(k_i)
  \nonumber \\
 & &  \mbox{} \times \exp
 \left[ i (k_p-k_s-k_i)z - i (\omega_{k_p}-\omega_{k_s}-
 \omega_{k_i}) t \right]
 + \mbox{H.c.} \nonumber \\
 &=& \hat{H}^{(-)}_{\rm int}(t) + \hat{H}^{(+)}_{\rm int}(t) .
\end{eqnarray}
where $ C_{\rm int} $ is a constant and $ \chi^{(2)} $ stands for
the second-order susceptibility.
The symbol $ {\cal E}_p^{(+)}(0,\omega_{k_p}-\omega^0_p) $ denotes
the positive-frequency
part of the envelope of the pump-beam electric-field amplitude
at the output plane
of the crystal; $ k_p $ stands for the wave vector of a mode
in the pump beam and $ \omega^0_p $ means the central frequency
of the pump beam. The symbol $ \hat{a}_s^{\dagger}(k_s) $
($ \hat{a}_i^{\dagger}(k_i) $) represents the creation
operator of the signal (idler) mode with wave vector $ k_s $
($ k_i $) and frequency $ \omega_{k_s} $ ($ \omega_{k_i} $).
The nonlinear crystal extends from $ z=-L $ to $ z=0 $. The
symbol $ {\rm H.c.} $ denotes Hermitian conjugate.
The operator $ \hat{H}^{(-)}_{\rm int} $
($ \hat{H}^{(+)}_{\rm int} $) stands for
the part of the interaction Hamiltonian $ \hat{H}_{\rm int} $
containing creation (annihilation) operators of modes in the
signal and idler fields.

The state of the signal and idler fields at the output
plane of the crystal determined by the solution of the
Schr\"{o}dinger equation can be written as follows:
\begin{eqnarray}      
 |\psi \rangle &=& \sum_{n=0}^{\infty} |\psi_n \rangle ,
   \nonumber \\
  |\psi_0 \rangle &=& |{\rm vac}\rangle , \nonumber \\
  |\psi_n \rangle &=& \left( - \frac{i}{\hbar} \right)^n
    \int_{-\infty}^{\infty} d\tau_1 \int_{-\infty}^{\tau_1}
    d\tau_2 \ldots \int_{-\infty}^{\tau_{n-1}} d\tau_n
    \nonumber \\
    & & \mbox{} \times
   \hat{H}_{\rm int}(\tau_1) \ldots \hat{H}_{\rm int}(\tau_n)
    |{\rm vac}\rangle , \hspace{1cm} n=1,2,\ldots .
\end{eqnarray}
We have assumed that the signal and idler fields
are in the vacuum state $ |{\rm vac}\rangle $ at the input
plane of the crystal.

Assuming that the number of photons in the signal and idler
fields is much lower than the number of modes constituting
these fields, we may approximately write:
\begin{eqnarray}      
   |\psi_n \rangle &=& \left( - \frac{i}{\hbar} \right)^n
    \int_{-\infty}^{\infty} d\tau_1 \int_{-\infty}^{\tau_1}
    d\tau_2 \ldots \int_{-\infty}^{\tau_{n-1}} d\tau_n
    \hat{H}^{(-)}_{\rm int}(\tau_1) \ldots \hat{H}^{(-)}_{\rm int}(\tau_n)
    |{\rm vac}\rangle \nonumber \\
    &=& \left( - \frac{i}{\hbar} \right)^n \frac{1}{n!}
    \left[ \int_{-\infty}^{\infty} d\tau
    \hat{H}^{(-)}_{\rm int}(\tau) \right]^n |{\rm vac}\rangle
     , \hspace{1cm} n=1,2,\ldots .
\end{eqnarray}
State $ |\psi_n\rangle $ then describes the field with exactly
$ n $ pairs in the signal and idler fields.

Photon statistics in the signal field may be determined
from the averages of the normally-ordered operators $ \hat{N}^{(n)}_s $
for $ n=1,2,\ldots $;
\begin{equation}    
  \hat{N}^{(n)}_s(\tau_1, \ldots, \tau_n,\tau_n,\ldots,\tau_1) =
  \left[ \prod_{j=1}^{n} \hat{E}^{(+)}_s(\tau_j) \right]
  \left[ \prod_{j=1}^{n} \hat{E}^{(-)}_s(\tau_j) \right] .
\end{equation}
The symbol
$ \hat{E}^{(+)}_s(\tau) $ ($ \hat{E}^{(-)}_s(\tau) $)
stands for the positive- (negative-) frequency part of
the electric-field amplitude of the signal field:
\begin{equation}   
 \hat{E}^{(+)}_s(\tau) = \sum_{k_s} e_s(k_s) \hat{a}_s(k_s)
 \exp (-i\omega_{k_s}\tau) .
\end{equation}
The symbol $ e_s(k_s) $ denotes the normalization amplitude
of the mode $ k_s $.

If the down-converted field is in the state $ |\psi_n\rangle $
given in Eq.~(B3), it holds for $ n \ge 1 $:
\begin{equation}     
  \langle \psi_n| \hat{N}^{(n)}_s(\tau_1,\ldots,\tau_n,
   \tau_n, \ldots,\tau_1)
  |\psi_n\rangle = {\cal P} \left\{ \prod_{j=1}^{n}
  \langle \psi_1| \hat{N}^{(1)}(\tau_{i_j},\tau_j) |\psi_1\rangle
  \right\} .
\end{equation}
The symbol $ {\cal P} $ means summation over all permutations
of the indices $ (i_1,\ldots,i_n) $ from the set $ (1,\ldots,n) $.
Assuming $ \langle \psi_n| \hat{N}^{(n)} |\psi_n\rangle
\gg \langle \psi_k| \hat{N}^{(n)} |\psi_k\rangle $ for
$ k=n+1, n+2 ,\ldots $, the relation in Eq.~(B6) implies
that photon statistics in the signal field is described by
the Bose-Einstein distribution.

In order to determine statistics of photon pairs,
we define the following ``creation operator of photon pairs'':
\begin{equation}   
 \hat{P}_{\rm pair}(\tau_s,\tau_i) =
 \underline{\hat{E}^{(-)}_s(\tau_{s}) \hat{E}^{(-)}_i(\tau_{i})} .
\end{equation}
Underlining of the operators on the right-hand side of Eq.~(B7)
means that ``only the signal and idler photons created in the
same elementary event are considered'' (see the expression
for $ |\psi_n\rangle $ in Eq.~(B3)).

We may write in the framework of the above used approximation:
\begin{eqnarray}    
  \langle \psi_n | \left[ \prod_{j=1}^{n}
  \hat{P}_{\rm pair}^\dagger(\tau_{s_j},\tau_{i_j}) \right]
  \left[ \prod_{j=1}^{n} \hat{P}_{\rm pair}(\tau_{s_{n+1-j}},
  \tau_{i_{n+1-j}}) \right] |\psi_n\rangle = & & \nonumber \\
  \left[ \prod_{j=1}^{n} \langle \psi_1 |
  \hat{P}_{\rm pair}^\dagger(\tau_{s_j},\tau_{i_j})
  |{\rm vac}\rangle \right]
  \left[ \prod_{j=1}^{n} \langle {\rm vac}|
  \hat{P}_{\rm pair}(\tau_{s_{n+1-j}},\tau_{i_{n+1-j}})
  |\psi_1 \rangle \right] , & & \nonumber \\
    n=1,2,\ldots . & &
\end{eqnarray}
Assuming that the contribution from the state $ |\psi_n\rangle $
is much greater that those from the states $ |\psi_k\rangle $
for $ k=n+1,n+2, \ldots $, the relation in Eq.~(B8) leads
to the conclusion that the statistics of photon pairs is
determined by the Poissonian distribution.

\section{Experimental determination of values of parameters occurring in the
model}

We give a connection of the model parameters $\mu$,
$T_S$, $T_I$, $\mu_S^{\rm res}$, and $\mu_I^{\rm res}$ to
the measured quantities. An experiment providing
detection rates $ n_S $ and $ n_I $
at detectors placed in the signal and idler beams and
coincidence-count rate $ n_c $ is assumed.

From the point of view of a real experimental setup,
the quantity $ \mu $ is determined by the number of photon pairs
beyond the nonlinear crystal such that at least one photon
of the pair has a nonzero probability of reaching a detector.
We further assume that $ \mu=k P $, where $ P $ is
the pump-laser power and $ k $ is an unknown constant.
We first describe the loss of photons caused by spatial
filtering of the signal and idler fields. The loss
is caused by the geometric placement of the detectors or pinholes or
fiber-coupling optics, whichever is the most limiting. We denote
the rate of pairs whose idler (signal) photon is absorbed [the
photon cannot reach a detector owing to spatial filtering] by
$ f_S \mu $ ($ f_I \mu $). The number of entangled pairs
in front of the detectors is then given by $ (1-f_S-f_I) \mu $.
The photons may also be lost owing to absorption and reflection
on their paths leading to the detectors (e.g.,
due to frequency filters). These losses are in general
different for `pairs' and `singles'. However, we consider them to be
the same and represent their influence by beamsplitters with
transmission coefficients $ t_S $ and $ t_I $ in the signal and idler
beams, respectively. This assumption is approximately valid when the
losses are only weakly spectrally dependent.

The coincidence-count rate $ n_c $ is written as:
\begin{equation}       
 n_c= d_S d_I + \mu t_S t_I (1-f_S-f_I) \eta_S \eta_I +
 O(\mu^2, d_S \mu, d_I \mu),
\end{equation}
where $ d_S $ ($d_I$) represents the dark-count rate and
$ \eta_S $ ($ \eta_I $) is the quantum efficiency of the detector
in the signal (idler) beam. This formula is
valid, e.g., when cw pumping of the process is applied
($ d_S, d_I \ll \mu \ll 1 $). Similarly, the detection rates in the
signal ($ n_S $) and idler ($ n_I $) beams are given as:
\begin{eqnarray}      
 n_S &=& d_S + \mu (1-f_I) t_S \eta_S +O(d_S \mu), \nonumber \\
 n_I &=& d_I + \mu (1-f_S) t_I \eta_I +O(d_I \mu).
\end{eqnarray}

Five unknown parameters $\mu$, $f_S$,
$f_I$, $t_S$, and $t_I$ cannot be uniquely determined from
Eqs.~(C1) and (C2).
In order to simplify the description, we first introduce the quantities
$ T_S $ [$T_S=t_S (1-f_I)$] and $ T_I $ [$T_I=t_I (1-f_S)$].
We then replace the coincidence-count rate $ n_c $ by
the quantity $ \tilde{n}_c $:
\begin{equation}      
 \tilde{n}_c= d_S d_I + \mu T_S T_I \eta_S \eta_I +
 O(\mu^2, d_S \mu, d_I \mu).
\end{equation}
The difference $ \tilde{n}_c - n_c $ equals to $\mu t_S
t_I f_S f_I \eta_S \eta_I $ and can be omitted if
$ f_S f_I \ll 1 $.

The dependencies of $ n_c $, $ n_S $, and $ n_I $ on the
pump-laser power $ P $ have been measured in the
experiment\footnote{Type-I nonlinear crystal has
been pumped using 0-420~mW of 413.1~nm line from a krypton-ion laser.
Correlated photon pairs have
been selected by pinholes and 5~nm (FWHM) interference filters and
then coupled to single-mode fibers that led them to silicon
avalanche photodetectors.} and the constants $ b_c $, $ b_S $, and
$ b_I $ characterizing the presumed linear dependencies on
the pumping power $ P $ have been
found. Eqs.~(C2) and (C3) then provide equations for the
determination of the parameters $ k $, $ T_S $, and $ T_I $:
\begin{eqnarray}     
 k T_S \eta_S &=& b_S, \nonumber \\
 k T_I \eta_I &=& b_I, \nonumber \\
 k T_S T_I \eta_S \eta_I &=& b_c.
\end{eqnarray}
Solving Eqs.~(C4), we have:
\begin{equation}     
T_S=\frac{b_c}{\eta_S b_I},\hspace{5mm} T_I=\frac{b_c}{\eta_I b_S},
\hspace{5mm} k=\frac{b_S b_I}{b_c} .
\end{equation}

We cannot determine the values of parameters $t_S$, $t_I$, $f_S$, and
$f_I$ in our experiment because it does not allow to
resolve two above discussed mechanisms causing decorrelation of
photons in a pair. However, we
can obtain limitations on their values taking into account
the relations $ 0 \leq t_S,t_I, f_S, f_I \leq 1 $:
\begin{eqnarray}          
T_S \leq t_S \leq 1, &\hspace{5mm} & T_I \leq t_I \leq 1, \nonumber \\
 0 \leq f_S \leq 1-T_I, &\hspace{5mm} & 0 \leq f_I \leq 1-T_S .
\end{eqnarray}
The knowledge of components of the
experimental setup may result in stronger limitations on
the values of the parameters $ t_S $ and $ t_I $ and
subsequently also on the values of $ f_S $ and $ f_I $.

If the assumption $ f_S f_I \ll 1 $ is not valid,
correct ratioes of the correlated and decorrelated photons
(given by $ n_c / n_S $, $ n_c / n_I $) may be kept
by introducing nonzero mean photon numbers of additional noisy fields
$ \mu^{\rm res, add}_S $ and $ \mu^{\rm res, add}_I $
[see Eqs.~(17) and (19);
$ n_c/n_S = \tilde{n}_c/(n_S + \mu^{\rm res, add}_S) $,
$ n_c/n_I = \tilde{n}_c/(n_I + \mu^{\rm res, add}_I) $]:
\begin{eqnarray}       
 \mu_S^{\rm res, add} &=& t_S f_S f_I \frac{1-f_I}{1-f_S-f_I} \mu,
 \nonumber \\
 \mu_I^{\rm res, add} &=& t_I f_S f_I \frac{1-f_S}{1-f_S-f_I}
 \mu.
\end{eqnarray}

As an example, we had $ \eta_S=0.474 $ and
$ \eta_I=0.586 $ in our setup and we measured
$ b_S=(5.13\pm0.05)\times10^{-5}$~W$ ^{-1} $,
$b_I=(5.50\pm0.04)\times10^{-5}$~W$ ^{-1} $, and
$b_c=(4.86\pm0.05)\times10^{-6}$~W$ ^{-1} $
(detection interval $\tau=1$~ns was used).
Using Eqs.~(C5), we arrive at:
\begin{eqnarray}     
 T_S &=& 0.186\pm0.002,\nonumber \\
 T_I &=& 0.162\pm0.002,\nonumber \\
  k &=& (5.81\pm0.09)\times10^{-4}~{\rm W}^{-1} .
\end{eqnarray}
Eqs.~(C6) provide the following limitations:
$$
  0 < f_S < 0.838,\hspace{0.5cm}  0 < f_I < 0.814 .
$$
Using the knowledge of components in the experimental setup,
we have $ t_S < 0.25 $ and $ t_I < 0.25 $ and subsequently
\begin{equation}    
 0 < f_S < 0.35,\hspace{5mm} 0 < f_I < 0.25.
\end{equation}
Values of the additional-noise terms are then bounded by
the inequalities $ 0 < \mu_S^{\rm res, add} < 0.041 \mu $ and
$ 0 < \mu_I^{\rm res, add} < 0.036 \mu $.

\section{Narrow spectra of the down-converted photons}

If the spectra of the down-converted photons are narrow,
pairs of photons obey the Bose-Einstein distribution
\cite{milburnwalls} and we have:
\begin{equation}     
 |c_n|^2 = (1-\nu) \nu^n , \hspace{1cm}
 \nu = \frac{\mu}{1+\mu} ,
\end{equation}
where $ \mu $ denotes the mean number of photon pairs.

Assuming chaotic noisy fields in the signal and
idler beams as given in Eq.~(13) and substituting
the expression in Eq.~(D1) into Eqs.~(4) and (12), the diagonal
matrix elements of the
statistical operator $ \hat{\rho}_{I,k}^{\rm mix} $
are determined as follows:
\begin{eqnarray}    
 (\hat{\rho}_{I,k}^{\rm mix})_{nn} &=&
 \frac{ (1- \nu^{\rm res}_{I,k}) (1-\nu {\cal B})
 (1-\nu {\cal A}_k) }{
  (1-\nu {\cal B}) - (1-d_k)(1-\nu{\cal A}_k) }
  \nonumber \\
  & & \mbox{} \times \left\{ \frac{1}{1-\nu R_I {\cal A}_k}
   \, g_n\left(\nu^{\rm res}_{I,k}, \frac{\nu T_I {\cal A}_k}{
   1-\nu R_I {\cal A}_k} \right) \right. \nonumber \\
  & & \mbox{} \hspace{5mm} \left. -
   \frac{1-d_k}{1-\nu R_I {\cal B}}
   \, g_n\left(\nu^{\rm res}_{I,k}, \frac{\nu T_I {\cal B}}{
   1-\nu R_I {\cal B}} \right) \right\} ,
\end{eqnarray}
and
\begin{equation}    
 g_n(x,y) = \frac{x^{n+1} - y^{n+1}}{ x-y } .
\end{equation}

The expressions for $ d^{\rm noise}_j $ in Eq.~(18) and
$ \mu^{\rm res}_{I,k} $ in Eq.~(19) remain valid also for
the coefficients $ c_n $ given in Eq.~(D1).
The quantity $ r_{I,k} $ is given according to the
relation:
\begin{equation}  
 r_{I,k} = \left[ \prod_{l=1,\ldots,N; l\ne k} (1-d_l) \right]
  \frac{1-\nu}{1-\nu{\cal A}_k}
  - \left[ \prod_{l=1}^{N} (1-d_l) \right]
   \frac{1-\nu}{1-\nu{\cal B}} .
\end{equation}

Assuming a symmetric $ 1 \times N $ coupler
(described in Eq.~(20)) and detection of a photon
at an arbitrary detector, we get:
\begin{eqnarray}    
 (\hat{\rho}_{I}^{\rm mix,s})_{nn} &=&
 \frac{ (1- \nu^{\rm res}_{I}) (1-\nu {\cal B})
 (1-\nu {\cal A}) }{
  (1-\nu {\cal B}) - (1-d)(1-\nu{\cal A}) }
  \nonumber \\
  & & \mbox{} \times \left\{ \frac{1}{1-\nu R_I {\cal A}}
   \, g_n\left(\nu^{\rm res}_{I}, \frac{\nu T_I {\cal A}}{
   1-\nu R_I {\cal A}} \right) \right. \nonumber \\
  & & \mbox{} \hspace{5mm} \left. -
   \frac{1-d}{1-\nu R_I {\cal B}}
   \, g_n\left(\nu^{\rm res}_{I}, \frac{\nu T_I {\cal B}}{
   1-\nu R_I {\cal B}} \right) \right\} ,
\end{eqnarray}
and
\begin{equation}  
 r_{I} = (1-d)^{N-1}
  \frac{1-\nu}{1-\nu{\cal A}}
  - (1-d)^N \frac{1-\nu}{1-\nu{\cal B}} .
\end{equation}

\end{document}